 \documentclass[preprint,aps]{revtex4}
 \usepackage{rotating}
 \newcommand{\SO}{\mathrm{SO}}

 \usepackage{amsmath}

\usepackage{graphicx}
 
\begin{document}
 \newcommand{\Qed}{\rule{2.5mm}{3mm}}
 \newcommand{\balpha}{\mbox{\boldmath {$\alpha$}}}
 \draft 
%
%
\title[On the origin of families of quarks and leptons]{On the origin of families of quarks 
and leptons - predictions for four families} 
\author{ G. Bregar, M. Breskvar, D. Lukman, 
N.S. Manko\v c Bor\v stnik\\
Department of Physics, University of
Ljubljana, Jadranska 19, 1000 Ljubljana, Slovenia\\
}
\date{\today}
\begin{abstract} 
The approach unifying all the internal degrees of freedom---proposed by one of 
us
---is offering a new way of understanding 
families of quarks and leptons:  
A part of the starting 
Lagrange density  in $d \;(=1+13)$, which includes two kinds of spin connection fields---the 
gauge fields of two types of Clifford algebra objects%
---transforms the right handed 
quarks and leptons into the left handed ones manifesting in $d=1+3$ the Yukawa couplings  
of the Standard model. 
We study the influence of the way of breaking symmetries on the
Yukawa couplings and estimate properties of the fourth family---the quark masses and the 
mixing matrix,  
investigating the  possibility that the fourth family of quarks and leptons appears at 
low enough energies to be observable with the new generation of accelerators.
\end{abstract}

\pacs{11.10.Kk,11.25.Mj,12.10.Dm,12.15.Ff,12.15.Hh}

\maketitle

\section{Introduction}
\label{introduction}

The Standard model of the electroweak and strong interactions (extended by assuming nonzero masses 
of the neutrinos) fits with around 25 parameters and constraints all the existing experimental 
data. It leaves, however,  unanswered many open questions, among which are also the questions   
about the origin of families, the Yukawa couplings of quarks and leptons and the 
corresponding Higgs mechanism.  
Understanding the mechanism for generating families, their masses and mixing matrices 
might be one of the most promising ways to physics beyond the Standard model.

The approach unifying spins and charges\cite{norma92,norma93,normasuper94,norma95,%
pikanormaproceedings1,holgernorma00,norma01,pikanormaproceedings2,Portoroz03,pikanorma05} 
might---by offering a new way of describing families---give an explanation 
about the origin of the Yukawa couplings. 
It was demonstrated in~\cite{pikanormaproceedings1,%
norma01,pikanormaproceedings2,Portoroz03} that a left handed $SO(1,13)$ 
Weyl spinor multiplet includes, if the representation is analyzed 
in terms of the subgroups $SO(1,3)$, $SU(2)$, $SU(3)$ and the sum of the two $U(1)$'s,  all 
the spinors of the Standard model---that is the left handed $SU(2)$ doublets and the right 
handed  $SU(2)$ singlets of quarks and leptons.
There are the (two kinds of)  spin connection  fields and  the vielbein fields in 
$d=(1+13)-$dimensional space, which might  manifest---after some  
appropriate compactifications (or some other kind of making the rest of $d-4$ space 
unobservable at low energies)---in the four dimensional 
space as all the gauge fields of the known charges, as well as the Yukawa couplings.

The paper \cite{pikanorma05}  analyzes, how do terms, which lead to masses of quarks 
and leptons, appear in the approach unifying spins and charges as a part of the spin 
connection and vielbein fields. No Higgs is needed in this approach to ''dress'' 
the right handed spinors with the weak charge, since the terms of the 
starting Lagrangean, which include $\gamma^0\gamma^s,$ with $s=7,8,$ do the job of a Higgs 
field. The approach predicts more than three families.

In this paper we study properties of mass matrices (Yukawa couplings) following from the 
approach unifying spins and charges after assuming some (two) possible breaks of symmetries. 
To calculate from the starting Lagrangean with only one (or at most two parameters) 
all the properties of the observed quarks and leptons is a too ambitious project 
at this moment (even in the limit 
when gravity can be treated---as it can be in this particular case, since breaks are supposed 
to occut far bellow the Planck scale---as ordinary 
gauge fields) because of all 
possible perturbative and nonperturbative effects. 
Instead we study the influence of particular breaks of symmetries 
from the  symmetry of the Lagrange density on properties of mass matrices. 
We make calculations on a tree level and leave the fields determining mass matrices 
after the assumed breaks as free  
parameters to be determined by the experimental data. In this way we try 
to understand how might the right way of breaking symmetries go if 
our approach has some meaning, and whether the fourth family of quarks and leptons 
might at all appear at 
energies observable with new accelerators. We present masses and mixing matrices for 
the four families of quarks. 

There are several attempts in the literature to explain the origin of families.  
In ref.~\cite{wz}, for example, 
the authors investigate the possibility that the unification of the charges (described by 
$SO(10)$) and the family quantum number (described by $SO(8)$) within   the group $SO(18)$ 
might be the right way to understand the replication of the family 
of quarks and leptons at low energies. 
In ref.~\cite{hld} the authors assume 
the Standard model group to the third power to reproduce families. 
In the approach unifying spins and charges 
spinors (due to two types of $\gamma^a$ operators) carry two indices, one  
index takes care of the ordinary spin, the other of the 
family. There are the ordinary $S^{ab}$ which transform one state of a spinor 
representation into another state, while $\tilde{S}^{ab}$ transform the family index. 

The approach unifying spins and charges shares with the Kaluza-Klein-like theories 
the difficulty how to ensure   
masslessness of  spinors in $d=1+3$ as well as their   
chiral coupling to the corresponding gauge fields (\cite{witten}). The  
Kaluza-Klein-like theories are also in danger to manifest in 
$d=1+3$  charges of both signs, in disagreement with the experimental data, since 
in the second quantization procedure 
spinors of the opposite charges (antiparticles) appear anyhow. 
We proposed in the ref. \cite{holgernorma05,hn07} the   
boundary condition for spinors in $d= 1+5$ compactified on a finite disk that ensures  
masslessness of  spinors in $d=1+3$ allowing at the same time charges of only one sign. 
We hope that such a toy model might be extended to the case of $d=1+13$.

Although we are far from being able to calculate from the simple starting action in $d=1+13$ 
the properties of the families of quarks and leptons as manifested at measurable energies 
directly---each break causes perturbative and nonperturbative effects, which are by 
themselves hard problems (not yet solved even in the hadron physics)---the approach 
manifests several nice features:\\ 
i) In one Weyl representation in $d=1+13$ all the quarks and the leptons of one 
family appear, but only the left handed quarks and leptons are weak charged while 
the right handed ones are weak chargeless. \\
ii) The starting Lagrange density offers the mechanism for generating families 
by including two kinds of the Clifford algebra objects. \\
iii) It is a part of the starting Lagrange density in $d=1+13$ which transforms 
the right handed weak chargeless spinors into the left handed weak charged spinors  
manifesting the Yukawa couplings of the Standard model.  

 The assumed breaks of symmetries relate mass matrix elements 
quite strongly and make accordingly possible 
predictions about properties of the fourth family of quarks and leptons in dependence 
of a way of breaking symmetries---after 
connecting free parameters of mass matrices with the known experimental data.


\section{Action for chargeless Weyl spinors in $d= (1+13)$ and appearance of
families of quarks and leptons} 
\label{lagrangesec}
%

This section repeats briefly the approach unifying spins and charges as presented 
in~ref.\cite{pikanorma05}. 
We assume that only a  left handed Weyl spinor  in $(1+13)$-dimensional space exists. 
A spinor carries 
only the spin (no charges) and interacts accordingly with only the gauge gravitational 
fields---with the spin connections and the vielbeins. We assume two kinds of the 
Clifford algebra objects 
and allow accordingly two kinds of gauge fields\cite{norma92,norma93,normasuper94,norma95,%
pikanormaproceedings1,holgernorma00,norma01,pikanormaproceedings2,Portoroz03,pikanorma05}. 
One kind is the ordinary gauge field (gauging the Poincar\' e symmetry in $d=1+13$). 
The corresponding spin connection fields appear for spinors as  gauge fields of 
$S^{ab}$ (Eq.\ref{tildesab}) defined in terms of $\gamma^a$, which are the ordinary Dirac operators
\begin{eqnarray}
S^{ab} &=& \frac{1}{2} (\gamma^a \gamma^b-
\gamma^b \gamma^a),\nonumber\\
\{\gamma^a,\gamma^b\}_+ &=& 2\eta^{ab}.   
\label{sab}
\end{eqnarray}
These gauge fields manifest at ''physical energies'' as all the gauge fields of the Standard model, 
and they also contribute---by connecting the right handed weak chargeless quarks or 
leptons to the left handed weak charged 
partners within one family of spinors---to the diagonal terms of mass matrices.     

The second kind of gauge fields is in our approach responsible for the appearance of 
families of spinors and accordingly also 
for couplings among families, contributing to diagonal matrix elements as well. 
It might explain, together  with the first kind of the spin connection fields,     
the Yukawa couplings of the 
Standard model of the electroweak and colour interactions. 
The corresponding spin connection fields appear for spinors as  gauge fields of $\tilde{S}^{ab}$ 
\begin{eqnarray}
\tilde{S}^{ab} &=& \frac{1}{2} (\tilde{\gamma}^a \tilde{\gamma}^b-
\tilde{\gamma}^b \tilde{\gamma}^a),\nonumber\\
\{\tilde{\gamma}^a,\tilde{\gamma}^b\}_+ &=& 2\eta^{ab}, \quad 
\{\tilde{\gamma}^a,\gamma^b\}_+ = 0, \quad \{ S^{ab}, \tilde{S}^{cd}\}_-=0,   
\label{tildesab}
\end{eqnarray}
with  $\tilde{\gamma}^a $ as the second kind of the Clifford algebra objects
\cite{norma93,technique03}.

Following the ref. \cite{pikanorma05} we write the action for a Weyl (massless) spinor  
in $d(=1+13)$-dimensional space as follows \footnote{Latin indices  
$a,b,..,m,n,..,s,t,..$ denote a tangent space (a flat index),
while Greek indices $\alpha, \beta,..,\mu, \nu,.. \sigma,\tau ..$ denote an Einstein 
index (a curved index). Letters  from the beginning of both the alphabets
indicate a general index ($a,b,c,..$   and $\alpha, \beta, \gamma,.. $ ), 
from the middle of both the alphabets   
the observed dimensions $0,1,2,3$ ($m,n,..$ and $\mu,\nu,..$), indices from the bottom of 
the alphabets
indicate the compactified dimensions ($s,t,..$ and $\sigma,\tau,..$). We assume the signature 
$\eta^{ab} =
diag\{1,-1,-1,\cdots,-1\}$.
}
\begin{eqnarray}
S &=& \int \; d^dx \; {\mathcal L},  
\nonumber\\
{\mathcal L} &=& \frac{1}{2} (E\bar{\psi}\gamma^a p_{0a} \psi) + h.c. = \frac{1}{2} 
(E\bar{\psi} \gamma^a f^{\alpha}{}_a p_{0\alpha}\psi) + h.c.,
\nonumber\\
p_{0\alpha} &=& p_{\alpha} - \frac{1}{2}S^{ab} \omega_{ab\alpha} - \frac{1}{2}\tilde{S}^{ab} 
\tilde{\omega}_{ab\alpha}.
\label{lagrange}
\end{eqnarray}
Here $f^{\alpha}{}_a$ are  vielbeins (inverted to the gauge field of the generators of translations  
$e^{a}{}_{\alpha}$, $e^{a}{}_{\alpha} f^{\alpha}{}_{b} = \delta^{a}_{b}$,
$e^{a}{}_{\alpha} f^{\beta}{}_{a} = \delta_{\alpha}{}^{\beta}$),
with $E = \det(e^{a}{}_{\alpha})$, while  
$\omega_{ab\alpha}$ and $\tilde{\omega}_{ab\alpha} $ are the two kinds of the spin connection fields, 
the gauge 
fields of $S^{ab}$ and $\tilde{S}^{ab}$, respectively.
 (We kindly ask the reader to read about the properties 
of these two kinds of the Clifford algebra objects - $\gamma^a$ and $\tilde{\gamma}^a$  
and of the corresponding  $S^{ab}$ and $\tilde{S}^{ab}$ - and about our technique in the 
ref. \cite{pikanorma05} or the refs. \cite{holgernorma02,technique03}.) 

One Weyl spinor representation in $d=(1+13)$ with the spin as the only internal 
degree of freedom    
manifests, if analyzed in terms of the subgroups $SO(1,3) \times
U(1) \times SU(2) \times SU(3)$ in 
four-dimensional ''physical'' space  as the ordinary ($SO(1,3)$) spinor with all the known charges 
of one family of  the left handed weak charged and the right handed weak chargeless 
quarks and leptons of the Standard model. To manifest this we make a choice of $\tau^{Ai}
 = \sum_{s,t} \;c^{Ai}{ }_{st} \; S^{st}, $ with $c^{Ai}{ }_{st}$ chosen in such a way that 
  $\tau^{Ai}$ 
 fulfil the commutation relations of the $SU(3), SU(2)$ and $U(1)$ groups:
$\{\tau^{Ai}, \tau^{Bj}\}_- = i \delta^{AB} f^{Aijk} \tau^{Ak},$ 
with the structure constants $f^{Aijk}$ of the corresponding groups, 
where the index $A$ identifies the charge groups ($A=1$ denotes $SU(2)$ and the weak charge, 
$A=2$ denotes one 
of the two $U(1)$ groups---the one following from $SO(1,7)$---$A=3$ denotes  
the group $SU(3)$ and the colour cahrge and $A=4$ denotes the group $U(1)$ 
following from SO(6)) and index $i$ identifies 
the generators within one charge group.
We have: 
$
\tau^{11}: = \frac{1}{2} ( {\mathcal S}^{58} - {\mathcal S}^{67} ),
\tau^{12}: = \frac{1}{2} ( {\mathcal S}^{57} + {\mathcal S}^{68} ),
\tau^{13}: = \frac{1}{2} ( {\mathcal S}^{56} - {\mathcal S}^{78} ),
\tau^{21}: = \frac{1}{2} ( {\mathcal S}^{56} + {\mathcal S}^{78} ),
\tau^{31}: = \frac{1}{2} ( {\mathcal S}^{9\;12} - {\mathcal S}^{10\;11} ),
\tau^{32}: = \frac{1}{2} ( {\mathcal S}^{9\;11} + {\mathcal S}^{10\;12} ),
\tau^{33}: = \frac{1}{2} ( {\mathcal S}^{9\;10} - {\mathcal S}^{11\;12} ),
\tau^{34}: = \frac{1}{2} ( {\mathcal S}^{9\;14} - {\mathcal S}^{10\;13} ),
\tau^{35}: = \frac{1}{2} ( {\mathcal S}^{9\;13} + {\mathcal S}^{10\;14} ),
\tau^{36}: = \frac{1}{2} ( {\mathcal S}^{11\;14} - {\mathcal S}^{12\;13}),
\tau^{37}: = \frac{1}{2} ( {\mathcal S}^{11\;13} + {\mathcal S}^{12\;14} ),
\tau^{38}: = \frac{1}{2\sqrt{3}} ( {\mathcal S}^{9\;10} + {\mathcal S}^{11\;12} - 2{\mathcal S}^{13\;14}),
\tau^{41}: = -\frac{1}{3}( {\mathcal S}^{9\;10} + {\mathcal S}^{11\;12} + {\mathcal S}^{13\;14}),$
and $Y = \tau^{41} + \tau^{21}, \quad  Y' = \tau^{41} - \tau^{21}.$
The reader can find this analyses in the paper \cite{pikanorma05}. 
We proceed as follows. We make a choice of the Cartan subalgebra set with $d/2=7$ elements 
in $d=1+13$: 
\begin{eqnarray}
\label{cartan}
S^{03}, S^{12}, S^{56}, 
S^{78}, S^{9\, 10}, S^{11 \,12}, S^{13 \,14}.
\end{eqnarray}
Then we express the basis for one Weyl in 
$d=1+13$ as products  of nilpotents and projectors  \cite{pikanorma05}
\begin{eqnarray}
\stackrel{ab}{(k)}: = \frac{1}{2} (\gamma^a + \frac{\eta^{aa}}{ik} \gamma^b ),\quad 
\stackrel{ab}{[k]}= \frac{1}{2} (1 + \frac{i}{k} \gamma^a \gamma^b ),  
\label{basis}
\end{eqnarray}
respectively, which  all are eigenvectors of $S^{ab}$ 
\begin{eqnarray}
S^{ab} \stackrel{ab}{(k)}: = \frac{k}{2} \stackrel{ab}{(k)},\quad 
S^{ab} \stackrel{ab}{[k]}: = \frac{k}{2} \stackrel{ab}{[k]}.  
\label{basis1}
\end{eqnarray}
We choose the starting vector to be an eigen vector of all the members of the Cartan set. 
In particular, the vector $\stackrel{03}{(+i)}\stackrel{12}{(+)}\stackrel{56}{(+)}\stackrel{78}{(+)}\;
\stackrel{9 \;10}{[-]}\;\stackrel{11\;12}{[+]}\; \stackrel{13\;14}{(-)}  
$ has the following eigenvalues of the Cartan subalgebra 
set:  
 $(\frac{i}{2},\frac{1}{2},\frac{1}{2},\frac{1}{2},-\frac{1}{2},
\frac{1}{2},-\frac{1}{2}),$ respectively. With respect to the charge groups it represents a right 
handed weak chargeless $u$-quark with spin up and with the 
colour $(-1/2, 1/(2\sqrt{3}))= (\tau^{33},\tau^{38})$. (How does the ordinary group 
theoretical way of analyzing spinors go can be found in many text books, 
also in ref.~\cite{wz}.)

Accordingly we may write 
one octet of  the left handed and the right handed 
quarks of both spins and of one colour charge as presented in Table I. 
\begin{table}
\begin{center}
\begin{tabular}{|r|c||c||c|c||c|c|c||c|c|c||r|r|}
\hline
i&$$&$|^a\psi_i>$&$\Gamma^{(1,3)}$&$ S^{12}$&$\Gamma^{(4)}$&
$\tau^{13}$&$\tau^{21}$&$\tau^{33}$&$\tau^{38}$&$\tau^{41}$&$Y$&$Y'$\\
\hline\hline
&& ${\rm Octet},\;\Gamma^{(1,7)} =1,\;\Gamma^{(6)} = -1,$&&&&&&&&&& \\
&& ${\rm of \; quarks}$&&&&&&&&&&\\
\hline\hline
1&$u_{R}^{c}$&$\stackrel{03}{(+i)}\stackrel{12}{(+)}|\stackrel{56}{(+)}\stackrel{78}{(+)}
||\stackrel{9 \;10}{[-]}\stackrel{11\;12}{[+]}\stackrel{13\;14}{(-)}$
&1&1/2&1&0&1/2&-1/2&$1/(2\sqrt{3})$&1/6&2/3&-1/3\\
\hline 
2&$u_{R}^{c}$&$\stackrel{03}{[-i]}\stackrel{12}{[-]}|\stackrel{56}{(+)}\stackrel{78}{(+)}
||\stackrel{9 \;10}{[-]}\stackrel{11\;12}{[+]}\stackrel{13\;14}{(-)}$
&1&-1/2&1&0&1/2&-1/2&$1/(2\sqrt{3})$&1/6&2/3&-1/3\\
\hline
3&$d_{R}^{c}$&$\stackrel{03}{(+i)}\stackrel{12}{(+)}|\stackrel{56}{[-]}\stackrel{78}{[-]}
||\stackrel{9 \;10}{[-]}\stackrel{11\;12}{[+]}\stackrel{13\;14}{(-)}$
&1&1/2&1&0&-1/2&-1/2&$1/(2\sqrt{3})$&1/6&-1/3&2/3\\
\hline 
4&$d_{R}^{c}$&$\stackrel{03}{[-i]}\stackrel{12}{[-]}|\stackrel{56}{[-]}\stackrel{78}{[-]}
||\stackrel{9 \;10}{[-]}\stackrel{11\;12}{[+]}\stackrel{13\;14}{(-)}$
&1&-1/2&1&0&-1/2&-1/2&$1/(2\sqrt{3})$&1/6&-1/3&2/3\\
\hline
5&$d_{L}^{c}$&$\stackrel{03}{[-i]}\stackrel{12}{(+)}|\stackrel{56}{[-]}\stackrel{78}{(+)}
||\stackrel{9 \;10}{[-]}\stackrel{11\;12}{[+]}\stackrel{13\;14}{(-)}$
&-1&1/2&-1&-1/2&0&-1/2&$1/(2\sqrt{3})$&1/6&1/6&1/6\\
\hline
6&$d_{L}^{c}$&$\stackrel{03}{(+i)}\stackrel{12}{[-]}|\stackrel{56}{[-]}\stackrel{78}{(+)}
||\stackrel{9 \;10}{[-]}\stackrel{11\;12}{[+]}\stackrel{13\;14}{(-)}$
&-1&-1/2&-1&-1/2&0&-1/2&$1/(2\sqrt{3})$&1/6&1/6&1/6\\
\hline
7&$u_{L}^{c}$&$\stackrel{03}{[-i]}\stackrel{12}{(+)}|\stackrel{56}{(+)}\stackrel{78}{[-]}
||\stackrel{9 \;10}{[-]}\stackrel{11\;12}{[+]}\stackrel{13\;14}{(-)}$
&-1&1/2&-1&1/2&0&-1/2&$1/(2\sqrt{3})$&1/6&1/6&1/6\\
\hline
8&$u_{L}^{c}$&$\stackrel{03}{(+i)}\stackrel{12}{[-]}|\stackrel{56}{(+)}\stackrel{78}{[-]}
||\stackrel{9 \;10}{[-]}\stackrel{11\;12}{[+]}\stackrel{13\;14}{(-)}$
&-1&-1/2&-1&1/2&0&-1/2&$1/(2\sqrt{3})$&1/6&1/6&1/6\\
\hline\hline
\end{tabular}
\end{center}
\caption{\label{TableI.} The 8-plet of quarks---the members of $SO(1,7)$ subgroup, 
belonging to one Weyl left 
handed ($\Gamma^{(1,13)} = -1 = \Gamma^{(1,7)} \times \Gamma^{(6)}$) spinor representation of 
$SO(1,13)$. 
It contains the left handed weak charged quarks and the right handed weak chargeless quarks 
of a particular 
colour $(-1/2,1/(2\sqrt{3}))$. Here  $\Gamma^{(1,3)}$ defines the handedness in $(1+3)$ space, 
$ S^{12}$ defines the ordinary spin (which can also be read directly from the basic vector), 
$\tau^{13}$ defines the weak charge, $\tau^{21}$ defines the $U(1)$ charge from $SO(1,7)$, 
$\tau^{33}$ and 
$\tau^{38}$ define the colour charge and $\tau^{41}$ defines another $U(1)$ charge, 
which together with the 
first one defines $Y = \tau^{41} + \tau^{21} $ and $Y'=\tau^{41} - \tau^{21}$. The vectors 
are eigenvectors of all the members of the Cartan subalgebra set ($\{S^{03}, S^{12}, S^{56}, 
S^{78}, S^{9 10}, S^{11 12}, S^{13 14}\}$).
The reader can find the whole Weyl representation in ref.~\cite{Portoroz03}.}
\end{table}

\noindent
All the members of the octet of Table~\ref{TableI.} can be obtained from the 
first state by the application  
of $S^{ab}; (a,b)= (0,1,2,3,4,5,6,7,8)$. 

%
The operators of handedness 
are defined as follows $\Gamma^{(1,13)}$ = $ i 2^{7} \; S^{03} S^{12} S^{56}$
$ \cdots S^{13 \; 14}, $ $ 
\Gamma^{(1,3)}$=$  - i 2^2 S^{03} S^{12}, $ $
\Gamma^{(1,7)}$=$  - i2^{4}  S^{03} S^{12} S^{56} S^{78}, $ $
\Gamma^{(6)}$ =$ - 2^3 S^{9 \;10} S^{11\;12} S^{13 \; 14}, $ $
\Gamma^{(4)}$=$ $ $2^2 S^{56} S^{78}.$

Quarks of the other two color charges and the color chargeless leptons 
distinguish from this octet only in the part which 
determines the color charge and $\tau^{41}$ ($\tau^{41}=1/6$ for quarks and $
\tau^{41}=-1/2$ for leptons). (They can be obtained from the octet of Table~\ref{TableI.} 
by the application of $S^{ab};
		(a,b)= (9,10,11,12,13,14)$ on these states. In particular, 
		$S^{9\;13}$ transforms 
the right handed $u_{R}^{c}$-quark of the first column into the right handed 
weak chargeless neutrino of the 
same spin ($\stackrel{03}{(+i)}\stackrel{12}{(+)}|\stackrel{56}{(+)}\stackrel{78}{(+)}
||\stackrel{9 \;10}{(+)}\stackrel{11\;12}{[+]}\stackrel{13\;14}{[+]}$), while it has 
 $\tau^{41}=-1/2$ and accordingly  $Y=0,Y'=-1$.)  One notices that $ 2 \tau^{41} $ 
 measures  the baryon number of quarks, while $ - 2 \tau^{41} $ measures the 
 lepton number. Both are 
 conserved quantities with respect to the group $SO(1,7)$.

We can formally rewrite the Lagrangean of Eq.(\ref{lagrange}) so that it manifests the usual 
Lagrange density for spinors in $d=(1+3)$ of the Standard model of the electroweak and 
colour interactions before the Higgs field breaks the $SU(2)\times U(1)$ 
symmetry and that it manifests 
the Yukawa couplings as well
\begin{eqnarray}
{\mathcal L} &=& \bar{\psi}\gamma^{m} (p_{m}- \sum_{A,i}\; g^{A}\tau^{Ai} A^{Ai}_{m}) \psi 
+ \nonumber\\
& &  \sum_{s=7,8}\; 
\bar{\psi} \gamma^{s} p_{0s} \; \psi + {\rm the \;rest}.
\label{yukawa}
\end{eqnarray}
Here $p_{0s}= p_s - \frac{1}{2} S^{s't}\omega_{s'ts} -  
 \frac{1}{2} \tilde{S}^{s't}\tilde{\omega}_{s't s}; \;
 (s',t) \in (5,6,..,)$, while $A^{Ai}_{m}, m=0,1,2,3,$ 
denote the 
gauge fields (expressible in terms of $\omega_{st m}$) corresponding to the 
charges defined by the generators $\tau^{Ai}$. 
One easily sees from Table \ref{TableI.} that the operator  
$\gamma^0 \gamma^7$ (or $\gamma^0 \gamma^8$ or any superposition of these two operators)  
transforms  the right handed weak chargeless $u_{R}^{c}$ quark of the first row into the 
left handed weak charged $u_{L}^{c}$ quark of the same spin and the colour 
charge presented on the seventh row---which is just what the Higgs field together 
with $\gamma^0$ do in the Standard model. 
We assume that  breaks of the starting symmetry $SO(1,13)$ (the Poincar\' e 
symmetry in $d=1+13$ and the symmetry in the $\tilde{S}^{ab}$ sector) lead  first to 
$SO(1,7) \times SU(3) \times U(1)$, where all the spinors are massless, while further breaks 
 lead to  $\SO(1,3)\times U(1)\times SU(3)$,  
manifesting the observed  symmetries and the Yukawa couplings with the observed 
masses of quarks and leptons and the mixing matrices. 
If finding out how and  at which scales does the break of $SO(1,7)$  
to  $SO(1,3) \times SU(2) \times U(1)$  (possibly via $SO(1,3) \times SO(4)$)  
occur the approach could accordingly offer the 
explanation, why do we observe spinors carrying beside the spin also the weak, 
the electromagnetic and the colour charge, why does each of the charges  couple 
with a different coupling constant to the corresponding gauge fields, 
what are the ratios of these coupling constants,  
why   have we observed up to now three families, all three of different masses  
and at which energy scale does the next family occur.  

We are not yet able to answer all these questions. Assuming that particular two ways of 
breaking  symmetries 
could occur, we are in this paper trying to 
find out possible connections between  breaks of symmetries and the 
symmetries of the corresponding  Yukawa couplings and to predict accordingly 
what are properties of the fourth family in each of these two ways of breaking symmetries. 
The larger are the symmetries of the 
Yukawa couplings after the assumed breaks of symmetries the smaller is the number of  
free parameters in the Yukawa couplings following from our approach and the 
more predictive is our approach unifying spins and charges in this simple application of it.

The terms responsible for the 
Yukawa couplings in our approach can be rearranged to be written in 
terms of nilpotents $\stackrel{78}{(\pm)}$ as follows 
\begin{eqnarray}
\label{popm}
\gamma^s p_{0s} =  
\stackrel{78}{(+)} p_{0_{78}+} +  \stackrel{78}{(-)} p_{0_{78}-},
\end{eqnarray}
with $s=7,8$ and $p_{0_{st}\pm} = p_{0s}\mp i p_{0t}$ and that we can write 
\begin{eqnarray}
\label{omegapm}
\tilde{S}^{ab}= \frac{i}{2} [\stackrel{ac}{\tilde{(k)}} + 
\stackrel{ac}{\tilde{(-k)}}][\stackrel{bc}{\tilde{(k)}} + 
\stackrel{bc}{\tilde{(-k)}}]
\end{eqnarray}
 for any $c$. 
 We  can accordingly  rewrite 
$ - \sum_{(a,b) } \frac{1}{2} \stackrel{78}{(\pm)}\tilde{S}^{ab} \tilde{\omega}_{ab\pm} =
- \sum_{(ac),(bd), \;  k,l}\stackrel{78}{(\pm)}\stackrel{ac}{\tilde{(k)}}\stackrel{bd}{\tilde{(l)}} 
\; \tilde{A}^{kl}_{\pm} ((ac),(bd)),$  
with the pair $(a,b)$ in the first sum running over all the  indices which do not characterize  
the Cartan subalgebra, with $ a,b = 0,\dots, 8$,  while the two pairs $(ac)$ and $(bd)$ 
in the second sum denote only the Cartan subalgebra pairs
 (for $SO(1,7)$ we only have the pairs $(03), (12)$; $(03), (56)$ ;$(03), (78)$;
$(12),(56)$; $(12), (78)$; $(56),(78)$ ); $k$ and $l$ run over four 
possible values so that $k=\pm i$, if $(ac) = (03)$ 
and $k=\pm 1$ in all other cases, while $l=\pm 1$. 

Having the spinor basis written in terms of projectors and nilpotents (Table \ref{TableI.}) 
and knowing the relations of Eq.(\ref{raiselower})
it turns out that it is convenient to rewrite the  
mass term $L_Y=\sum_{s=7,8}\; \bar{\psi} \gamma^{s} p_{0s} \; \psi$ in Eq.(\ref{yukawa}) 
as follows

\begin{eqnarray}
{\mathcal L}_{Y} = \psi^+ \gamma^0 \;  
\{ & &\stackrel{78}{(+)} ( \sum_{y=\tau^{21},\tau^{41}}\; y A^{y}_{+} + 
\sum_{\tilde{y}=\tilde{N}^{3}_{+},\tilde{N}^{3}_{-},\tilde{\tau}^{13},
\tilde{\tau}^{21},\tilde{\tau}^{41}} 
\tilde{y} \tilde{A}^{\tilde{y}}_{+}\;)\; + \nonumber\\
  & & \stackrel{78}{(-)} ( \sum_{y=\tau^{21},\tau^{41}}\;y  A^{y}_{-} +  
\sum_{\tilde{y}= \tilde{N}^{3}_{+},\tilde{N}^{3}_{-},\tilde{\tau}^{13},
\tilde{\tau}^{21},\tilde{\tau}^{41}} 
\tilde{y} \tilde{A}^{\tilde{y}}_{-}\;) + \nonumber\\
 & & \stackrel{78}{(+)} \sum_{\{(ac)(bd) \},k,l} \; \stackrel{ac}{\tilde{(k)}} 
 \stackrel{bd}{\tilde{(l)}} \tilde{{A}}^{kl}_{+}((ac),(bd)) \;\;+  \nonumber\\
 & & \stackrel{78}{(-)} \sum_{\{(ac)(bd) \},k,l} \; \stackrel{ac}{\tilde{(k)}} 
 \stackrel{bd}{\tilde{(l)}} \tilde{{A}}^{kl}_{-}((ac),(bd))\}\psi.
\label{yukawatilde0}
\end{eqnarray}
Taking into account that $\stackrel{78}{(+)}\stackrel{78}{(+)}=0=
\stackrel{78}{(-)}\stackrel{78}{[-]}, $ 
while $\stackrel{78}{(+)}\stackrel{78}{[-]}=\stackrel{78}{(+)} $ and  
$\stackrel{78}{(-)}\stackrel{78}{(+)}=-\stackrel{78}{[-]}, $ we recognize that 
Eq.(\ref{yukawatilde0}) 
distinguishes between  the $u$-quark (only the terms with $\stackrel{78}{(-)}$ 
give nonzero contributions) and the  
$d$-quarks (only the terms with $\stackrel{78}{(+)}$ give nonzero contributions) 
and accordingly also 
between the neutrino and the electron. We also see that the first two rows contribute 
to the diagonal elements 
of the mass matrices, while the second two contribute to their non diagonal elements. 
Both, diagonal and non diagonal elements are expressible in terms of the gauge fields 
$\omega_{abc}$ and $\tilde{\omega}_{abc}$. 
The diagonal matrix elements are expressed as the gauge fields of the operators $\tau^{21},\tau^{41}$ 
as well as the operators 
$\tilde{N}^{3}_{\pm}: = \frac{1}{2} ( \tilde{{\mathcal S}}^{12} \pm i
\tilde{{\mathcal S}}^{03} ), \; 
\tilde{\tau}^{13} : = \frac{1}{2} ( \tilde{{\mathcal S}}^{56} - \tilde{{\mathcal S}}^{78} ), \;
\tilde{\tau}^{21}: = \frac{1}{2} ( \tilde{{\mathcal S}}^{56} + \tilde{{\mathcal S}}^{78} ), \;
\tilde{\tau}^{41}: = -\frac{1}{3} ( \tilde{{\mathcal S}}^{9 \;10} + \tilde{{\mathcal S}}^{11\; 12} 
+  \tilde{{\mathcal S}}^{13\; 14} ).
$
Taking into account that
$
-\frac{1}{2} S^{st} \omega_{st\pm} = \tau^{21} A^{21}_{\pm} + \tau^{41} A^{41}_{\pm}, $
$-\frac{1}{2} \tilde{S}^{st} \tilde{\omega}_{st\pm} = \tilde{\tau}^{21} \tilde{A}^{21}_{\pm} + 
\tilde{\tau}^{41} \tilde{A}^{41}_{\pm} + \tilde{\tau}^{13} \tilde{A}^{13}_{\pm}, \;
-\frac{1}{2} \tilde{S}^{mn}\tilde{\omega}_{mn\pm} = 
\tilde{N}^{3}_{+} \tilde{A}^{\tilde{N}^{3}_{+}}_{\pm} +
\tilde{N}^{3}_{-} \tilde{A}^{\tilde{N}^{3}_{-}}_{\pm},
$ 
with the pairs $(m,n) =(0,3),(1,2)$; $(s,t) = (5,6),(7,8),$ belonging to the Cartan subalgebra and 
$\Omega_{\pm} = \Omega_7 \mp i \Omega_8$, where $\Omega_7, \Omega_8$ 
mean any of the above fields $\tilde{\omega}_{ab7}, \tilde{\omega}_{ab8}$, we find
$
A^{13}_{\pm} = - (\omega_{56\pm} - \omega_{78 \pm}), \;
A^{21}_{\pm}  = -\frac{1}{2}  (\omega_{56\pm} + \omega_{78\pm}),\;
A^{41}_{\pm} = -\frac{1}{2} (\omega_{9\;10 \pm} + \omega_{11\;12 \pm} + 
\omega_{13\;14\pm}), \;$ and equivalently $
\tilde{A}^{13}_{\pm} = - (\tilde{\omega}_{56\pm} - \tilde{\omega}_{78 \pm}), \;
\tilde{A}^{21}_{\pm}  = -\frac{1}{2}  (\tilde{\omega}_{56\pm} + \tilde{\omega}_{78\pm}),\;
\tilde{A}^{41}_{\pm} = -\frac{1}{2} (\tilde{\omega}_{9\;10 \pm} + \tilde{\omega}_{11\;12 \pm} + 
\tilde{\omega}_{13\;14\pm}), \;$ 
$\tilde{A}^{\tilde{N}^{3}_{+}}_{\pm} = - (\tilde{\omega}_{12 \pm} - i \; \tilde{\omega}_{03\pm}),\;
\tilde{A}^{\tilde{N}^{3}_{-}}_{\pm} = - (\tilde{\omega}_{12 \pm} + i \; \tilde{\omega}_{03\pm}).$
Let us point out that this is true only before any break of the symmetries occur. 
We repeat that $\omega_{ab c} = 
f^{\alpha}{}_{c} \;\omega_{ab \alpha}$ and $\tilde{\omega}_{ab c} = f^{\alpha}{}_{c} \;
\tilde{\omega}_{ab \alpha}$.  

We have for the non diagonal mass matrix the elements 
\begin{eqnarray}
\tilde{A}^{++}_{\pm} ((ab),(cd)) &=& -\frac{i}{2} (\tilde{\omega}_{ac\pm} 
-\frac{i}{r} \tilde{\omega}_{bc\pm} 
-i \tilde{\omega}_{ad\pm} -\frac{1}{r} \tilde{\omega}_{bd\pm} ), \nonumber\\
\tilde{A}^{--}_{\pm} ((ab),(cd)) &=& -\frac{i}{2} (\tilde{\omega}_{ac\pm} 
+\frac{i}{r} \tilde{\omega}_{bc\pm} 
+i \tilde{\omega}_{ad\pm} -\frac{1}{r} \tilde{\omega}_{bd\pm} ),\nonumber\\
\tilde{A}^{-+}_{\pm} ((ab),(cd)) &=& -\frac{i}{2} (\tilde{\omega}_{ac\pm} 
+ \frac{i}{r} \tilde{\omega}_{bc\pm} 
-i  \tilde{\omega}_{ad\pm} +\frac{1}{r} \tilde{\omega}_{bd\pm} ), \nonumber\\
\tilde{A}^{+-}_{\pm} ((ab),(cd)) &=& -\frac{i}{2} (\tilde{\omega}_{ac\pm} 
- \frac{i}{r} \tilde{\omega}_{bc\pm} 
+i  \tilde{\omega}_{ad\pm} +\frac{1}{r} \tilde{\omega}_{bd\pm} ),
\label{Awithomega}
\end{eqnarray}
with $r=i$, if $(ab) = (03)$ and $r=1$ otherwise.
 We simplify the index $kl$ in the exponent 
of fields $\tilde{A}^{kl}{}_{\pm} ((ac),(bd))$ to $\pm $, omitting $i$.

A way of breaking  any of the two symmetries - the Poincar\' e one and 
the symmetry  determined by the generators $\tilde{S}^{ab}$ in $d=1+13$ - 
strongly influences the Yukawa couplings of 
Eq.(\ref{yukawatilde0}), relating  the parameters $\tilde{\omega}_{abc}$ and influencing 
the coupling constants.

In this paper we assume two ways of breaking symmetries and  investigate 
under which conditions each of these 
two ways of breaking symmetries leads to up to now measured properties of 
fermions.


%
\subsection{Properties of Clifford algebra objects}
\label{tau}

Since $S^{ab}= \frac{i}{2}\gamma^a \gamma^b,$ for $a\ne b$ (for $a=b$ $S^{ab}=0$), 
it is useful to know the following properties of $\gamma^a$'s, if they 
are applied on nilpotents and projectors  
\begin{eqnarray}
\label{gammanilpro}
\gamma^a \stackrel{ab}{(k)}&=&\eta^{aa}\stackrel{ab}{[-k]},\; \;\;
\gamma^b \stackrel{ab}{(k)}= -ik \stackrel{ab}{[-k]},\nonumber\\
\gamma^a \stackrel{ab}{[k]}&=& \stackrel{ab}{(-k)},\;\;\;\quad\;\;
\gamma^b \stackrel{ab}{[k]}= -ik \eta^{aa} \stackrel{ab}{(-k)}.
\end{eqnarray}
Accordingly,  for example, $S^{ac} \stackrel{ab}{(k_1)}\stackrel{cd}{(k_2)} =
-i \frac{1}{2}\eta^{aa}\eta^{cc}\stackrel{ab}{[-k_1]}\stackrel{cd}{[-k_2]}$. 
The operators, which are an even product of nilpotents
\begin{eqnarray}
\label{raiselower0}
\tau^{\pm}_{(ab,cd),k_1,k_2} =  \stackrel{ab}{(\pm k_1)}\stackrel{cd}{(\pm k_2)},  
\end{eqnarray}
appear to be the raising and lowering operators for a particular pair $(ab, cd)$ 
belonging to the Cartan subalgebra of the group $SO(q, d-q)$, with $q=1$ in our case. 
There are always four possibilities for products of nilpotents with respect to the 
sign of $(k_1)$ and $(k_2)$, since $k_l = \pm i, l=1,2$  or $k_l = \pm 1, l=1,2$ 
(whether we have 
$i$ or $1$ depends on the 
character of the indices of the Cartan subalgebra: $i$ for the pair $(03)$ and $1$ otherwise). 
We can make use of $R$ and $L$  instead of $k_1,k_2$ to distinguish 
between the two kinds of lowering 
and raising operators in $SO(1,7)$, respectively, since they distinguish 
between right handed weak chargeless states and left handed 
weak charged states: 
When applied on states of inappropriate handedness $\tau^{\pm}_{(ab,cd),k_1,k_2}$ gives $0$. 
For example, 
$\tau^{\pm}_{(03,12),R} =  \stackrel{03}{(\pm i)}\stackrel{12}{(\pm)}$ 
is the raising ($\stackrel{03}{(+i)}\stackrel{12}{(+)}$) and lowering 
($\stackrel{03}{(-i)}\stackrel{12}{(-)}$) operator, respectively, 
for a right handed quark or lepton, while 
$\tau^{\pm}_{(03,12),L} =  \mp \stackrel{03}{(\mp i)}\stackrel{12}{(\pm)}$ is the 
corresponding left handed raising and lowering operator, respectively for left 
handed quarks and leptons.  
Being applied on a weak chargeless $u^{c}_{R}$ of a colour $c$ and of the spin $1/2$, 
$\tau^{-}_{(03,12),R}$ transforms it  to a weak chargeless $u^{c}_{R}$ of the same colour 
and handedness but of the spin $-1/2$, 
while $\tau^{-}_{(03,78),R} =  \stackrel{03}{(-i)}\stackrel{78}{(-)}$ transforms  a weak chargeless 
$u^{c}_{R}$ of any colour and of the spin $1/2$ into the weak charged $u^{c}_{L}$ of the same 
colour and the same spin but of the opposite  handedness.

It is useful to have in mind~\cite{holgernorma02,technique03} the following properties of the nilpotents
 $\stackrel{ab}{(k)}$:
\begin{eqnarray}
\stackrel{ab}{(k)}\stackrel{ab}{(k)}& =& 0, \quad \quad \stackrel{ab}{(k)}\stackrel{ab}{(-k)}
= \eta^{aa}  \stackrel{ab}{[k]}, \quad 
\stackrel{ab}{[k]}\stackrel{ab}{[k]} =  \stackrel{ab}{[k]}, \quad \quad
\stackrel{ab}{[k]}\stackrel{ab}{[-k]}= 0,  \nonumber\\
\stackrel{ab}{(k)}\stackrel{ab}{[k]}& =& 0,\quad \quad \quad \stackrel{ab}{[k]}\stackrel{ab}{(k)}
=  \stackrel{ab}{(k)}, \quad \quad 
\stackrel{ab}{(k)}\stackrel{ab}{[-k]} =  \stackrel{ab}{(k)},
\quad \quad \stackrel{ab}{[k]}\stackrel{ab}{(-k)} =0,  
\label{raiselower}
\end{eqnarray}
which the reader can easily check if taking into account Eq.(\ref{gammanilpro}). 
\subsection{Families of spinors}
\label{families}

Commuting with $S^{ab}$ ($\{\tilde{S}^{ab},S^{ab} \}_-=0$), 
the generators $\tilde{S}^{ab}$ generate  
equivalent representations, which we  recognize as families.
To evaluate the application of $\tilde{S}^{ab}$  on the starting family,  presented in 
Table \ref{TableI.}, we  take into account the Clifford algebra properties of $\tilde{\gamma}^a$. 
We find
\begin{eqnarray}
\tilde{\gamma^a} \stackrel{ab}{(k)} &=& - i\eta^{aa}\stackrel{ab}{[k]},\quad\;\,
\tilde{\gamma^b} \stackrel{ab}{(k)} =  - k \stackrel{ab}{[k]}, \nonumber\\
\tilde{\gamma^a} \stackrel{ab}{[k]} &=&  \;\;i\stackrel{ab}{(k)},\quad \quad \quad
\tilde{\gamma^b} \stackrel{ab}{[k]} =  -k \eta^{aa} \stackrel{ab}{(k)}.
\label{gammatilde}
\end{eqnarray}
Accordingly it follows 
\begin{eqnarray}
\stackrel{ab}{\tilde{(k)}} \stackrel{ab}{(k)}& =& 0, 
\quad \quad \;\,\stackrel{ab}{\tilde{(-k)}} \stackrel{ab}{(k)}
= -i \eta^{aa}  \stackrel{ab}{[k]},\;\, 
\stackrel{ab}{\tilde{(-k)}}\stackrel{ab}{[-k]}= i \stackrel{ab}{(-k)},\quad
\stackrel{ab}{\tilde{(k)}} \stackrel{ab}{[-k]} = 0, \nonumber\\
\stackrel{ab}{\tilde{(k)}} \stackrel{ab}{[k]}& =& i \stackrel{ab}{(k)}, \;\;
\stackrel{ab}{\tilde{(-k)}}\stackrel{ab}{[+k]}= 0, \;\;\quad \quad  \quad \stackrel{ab}{\tilde{(-k)}}\stackrel{ab}{(-k)}=0,
 \;\;\stackrel{ab}{\tilde{(k)}}\stackrel{ab}{(-k)} = -i \eta^{aa} \stackrel{ab}{[-k]}.
\label{raiselowertilde}
\end{eqnarray}

The operators, which are an even product of nilpotents in the $\tilde{\gamma}^a$ sector 
\begin{eqnarray}
\label{raiselowertilde1}
\tilde{\tau}^{\pm}_{(ab,cd),k_1,k_2} =  \stackrel{ab}{\tilde{(\pm k_1)}} 
\stackrel{cd}{\tilde{(\pm k_2)}},  
\end{eqnarray}
appear (equivalently as $\tau^{\pm}_{(ab,cd),k_1,k_2}$ in the $S^{ac}$ sector) 
as the raising and lowering operators, when a  pair $(ab),(cd)$ belongs to 
the Cartan subalgebra of the algebra $\tilde{S}^{ac}$,  
transforming a member of one family  into the same member of another family. 
For example: $\tilde{\tau}^{-}_{(03,12),-i,-1} = \tilde{\tau}^{-}_{(03,12),R} 
=  \stackrel{03}{\tilde{(-i)}}\stackrel{12}{\tilde{(-1)}}$ transforms the right handed 
$u_R^{c}$ quark from Table~\ref{TableI.} into the right handed 
$u_R^{c}$ quark $u_{R}^{c}=\stackrel{03}{[+i]}\stackrel{12}{[+]}|\stackrel{56}{(+)}\stackrel{78}{(+)}
||\stackrel{9 \;10}{[-]}\stackrel{11\;12}{[+]}\stackrel{13\;14}{(-)}$, which has all the properties 
with respect to the operators $S^{ab}$ the same as  $u_{R}^{c}$ from Table~\ref{TableI.}.

\section{From eight to four families of quarks and leptons}
\label{Lwithassumptions}

Assuming that the break of the symmetry from $SO(1,13)$  to $SO(1,7)\times SU(3)\times U(1)$ 
makes all the families except the massless ones determined by $SO(1,7)$ very heavy (of 
the order of $10^{15}\,$GeV or heavier), we and up with eight families:  
$\tilde{S}^{ab},$ with $(a,b)\in\{0,1,2,3,5,6,7,8\}, $ 
(or equivalently the products of nilpotents $\stackrel{ab}{\tilde{(k_1)}}
\stackrel{cd}{\tilde{(k_2)}}$, with $k_1, k_2$ equal to $\pm 1$ or $\pm i$, while 
$(ab), (cd)$ denote two of the four Cartan subalgebra pairs \{(03), (12), (56), (78)\}) 
generate $2^{8/2-1}$ families. 
The first member of the $SO(1,7)$ multiplet in Table~\ref{TableI.} (the right handed weak chargeless 
$u_{R}^c$-quark with spin $1/2$, for example, as well as the right handed weak chargeless 
neutrino with spin $1/2$---both differ 
only in the part which concerns the $SU(3)$ and the  $U(1)$ charge ($U(1)$ from $SO(6)$) and  
which stay unchanged under the application of $\tilde{S}^{ab}, $ 
with $(a,b)\in\{0,1,2,3,5,6,7,8\},$) appears in the following  
 $8$ families 
\begin{eqnarray}
{\rm I.}\;\;& & \stackrel{03}{(+i)} \stackrel{12}{(+)} |\stackrel{56}{(+)} \stackrel{78}{(+)}||\cdots \quad 
\;\;{\rm V.}\;\; \stackrel{03}{[+i]} \stackrel{12}{(+)} |\stackrel{56}{(+)} \stackrel{78}{[+]}||\cdots \nonumber\\
{\rm II.}\;\;& &\stackrel{03}{[+i]} \stackrel{12}{[+]} |\stackrel{56}{(+)} \stackrel{78}{(+)}||\cdots \quad
\;\;{\rm VI.}\;\; \stackrel{03}{(+i)} \stackrel{12}{[+]} |\stackrel{56}{[+]} \stackrel{78}{(+)}||\cdots \nonumber\\
{\rm III.}\;\;& &\stackrel{03}{(+i)} \stackrel{12}{(+)} |\stackrel{56}{[+]} \stackrel{78}{[+]}||\cdots\quad
{\rm VII.}\;\; \stackrel{03}{[+i]} \stackrel{12}{(+)} |\stackrel{56}{[+]} \stackrel{78}{(+)}||\cdots \nonumber\\
{\rm IV.}\;\;& &\stackrel{03}{[+i]} \stackrel{12}{[+]} |\stackrel{56}{[+]} \stackrel{78}{[+]}||\cdots\quad 
{\rm VIII.}\;\;\stackrel{03}{(+i)} \stackrel{12}{[+]} |\stackrel{56}{(+)} \stackrel{78}{[+]}||\cdots \;.
\label{eightfamilies}
\end{eqnarray}
The rest seven members of each of the above eight families can be obtained, as in Table \ref{TableI.}, 
by the application 
of the operators $S^{ab}$ on the above particular member (or with the help of 
the raising and lowering operators $\tau^{\pm}_{(ab,cd),k_1,k_2} $). 
One easily checks (by checking the 
quantum numbers represented in Table~\ref{TableI.}) that each of the eight states of 
Eq.(\ref{eightfamilies}) 
represents indeed the right handed weak chargeless quark (or the right handed weak chargeless  
lepton, depending on what appears  for $|| \cdots $ in Eq.(\ref{eightfamilies})). 
%

A way of breaking  further the symmetry $SO(1,7)\times U(1) $ 
$\times SU(3)$ 
influences strongly properties of  the mass  matrix elements determined by Eq.(\ref{yukawatilde0}). 
We assume  two  particular 
ways of breaking the symmetry $SO(1,7)\times U(1)$ and study under which 
conditions can the two ways of breaking symmetries reproduce the known experimental data.

To come from the starting action of the proposed approach (with at most two free parameters) 
to  the effective  action manifesting the Standard model of the electroweak and colour 
interaction---in 
this paper we treat only the  Yukawa part of the Standard model action---and further to 
the observed families as well as to make predictions for the properties of a possible 
fourth  family, 
we make  the following assumptions:

i.  The break of symmetries of the group $SO(1,13)$ (the  Poincar\' e group in 
$d=1+13$)  into $SO(1,7)\times SU(3)\times U(1)$ occurs 
in a way that in $d=1+7$ massless spinors with the charge $ SU(3)\times U(1)$ appear.
(Our work on the compactification 
of a massless spinor in $d=1+5$ into   $d=1+3$ and a finite disk gives us some hope that such 
an assumption might be justified\cite{holgernorma05,hn07}.) 
The break of symmetries influences  both, the (Poincar\' e) symmetry 
described by $S^{ab}$ and the symmetries described by $\tilde{S}^{ab}$. 

ii. Further breaks lead to two (almost) decoupled massive four families, well separated in masses.

iii. We  make estimates on a ''tree level''.  

iv. We assume the mass matrices to  be real and symmetric expecting that the complexity and 
the nonsymmetric properties of the mass matrices do not influence considerably masses and 
the real part of 
the mixing matrices of quarks and leptons. 
In this paper we do not study the $CP $ breaking.

The following two ways of breaking symmetries leading to four ''low lying'' families 
of quarks and leptons are chosen: \\

\noindent
a. First we  assume that the break of symmetries from $ SO(1,7)\times U(1)\times SU(3)$ 
to the observed symmetries in the "low energy" regime occurs so that all the  
non diagonal elements in the Lagrange density (Eq.(\ref{yukawatilde0})) 
caused by  the  operators  of the type $\stackrel{ab}{(k)} \stackrel{cd}{(l)} $ or of the type 
$\stackrel{ab}{\tilde{(k)}} \stackrel{cd}{\tilde{(l)}} $, with either $(ab)$ or $(cd)$ equal to 
$(56)$, are zero. In the ''Poincar\' e'' sector this assumption guarantees 
the conservation of  the electromagnetic charge $Q= S^{56} + \tau^{41}$ by the mass term, since 
the operators  $\stackrel{ab}{(k)} \stackrel{cd}{(l)} $ transform the $u$-quark into the $d$ quark 
and opposite.  
We extend this requirement 
also to the operators $\stackrel{ab}{\tilde{(k)}} \stackrel{cd}{\tilde{(l)}} $. 
This means that all the fields of the type 
$\tilde{A}^{kl}_{\pm} ((ab),(cd))$,
with either $k$ or $l$ equal to $\pm$ and with either 
$(ab)$ or $(cd)$  equal to $(56)$, are put to zero. Then the eight families split into decoupled 
two times  four families. One easily sees that the diagonal matrix elements 
can be chosen in such a way 
that one of the two four families has  much larger diagonal elements  
then the other  (which guarantees correspondingly  also much higher masses of the 
corresponding fermions).  Accordingly we are left to study the properties of one four family, 
decoupled from the other four family.
We present this study in subsection~\ref{MDN4}.\\

\noindent
b. In the second way of breaking symmetries from $ SO(1,7)\times U(1)\times SU(3)$ to the 
observed ''low energy'' sector  
we assume that no matrix elements of the type $S^{ms}\omega_{msc}$ or 
$\tilde{S}^{sm}\tilde{\omega}_{smc}$, with $m=0,1,2,3,$ and $s-5,6,7,8,$ are allowed. This means 
that all the matrix elements of the type $\tilde{A}^{kl}_{\pm} ((ab),(cd))$, 
with   either $k$ or $l$ equal to $\pm$ and with 
$(ab)$ equal to $(03)$ or $(12)$ and  $(cd)$  equal to $(56)$ or $(78)$, are put to zero.   
This  means that the symmetry $SO(1,7)\times U(1)$ breaks into $SO(1,3)\times SO(4)\times U(1)$ 
and further  into $SO(1,3)\times U(1)$.  Again the mass matrix of eight families 
splits into two times  decoupled four families.  
We  recognize that in this way of breaking symmetries the diagonal matrix elements 
of the higher four 
families are again much larger than the diagonal matrix elements of the lower four families. 
We study the properties of the four families  with the lower diagonal 
matrix elements in subsection~\ref{GN4}.\\

To simplify the problem we assume in both cases, in  a. and in b.,  that the mass 
matrices are  real and symmetric. To determine free parameters of   
mass matrices by fitting masses and  mixing matrices of four families to the 
measured values for the three known families within the known accuracy,  
is by itself quite a demanding task.  And we hope that 
after analyzing two  possible  breaks of symmetries even   
such a simplified study can help to  understand the origin of families and  
to predict properties of the fourth family.

\subsection{Four families of quarks in proposal no. I}
\label{MDN4}

The assumption that there are no matrix elements  of the type 
$\tilde{A}^{k l}_{\pm} ((ab),(cd))$, 
with $k=\pm $ and $l=\pm $ (in all four combinations) and with either 
$(ab)$ or $(cd)$  equal to $(56)$ leads to the following four families (corresponding to
the families I,II,IV,VIII in Eq.(\ref{eightfamilies}))
\begin{eqnarray}
I.\;& & \stackrel{03}{(+i)} \stackrel{12}{(+)} |\stackrel{56}{(+)} \stackrel{78}{(+)}||...\nonumber\\ 
II.\;& &\stackrel{03}{[+i]} \stackrel{12}{[+]} |\stackrel{56}{(+)} \stackrel{78}{(+)}||... \nonumber\\
III.& & \stackrel{03}{[+i]} \stackrel{12}{(+)} |\stackrel{56}{(+)} \stackrel{78}{[+]}||... \nonumber\\
IV. & & \stackrel{03}{(+i)} \stackrel{12}{[+]} |\stackrel{56}{(+)} \stackrel{78}{[+]}||... .
\label{fourfamilies}
\end{eqnarray}
and to the corresponding mass matrices presented in Table \ref{TableII.} and Table \ref{TableIII.}. 
It is easy to see that the parameters can be chosen 
so that the second four families, decoupled from the first four, have much higher 
diagonal matrix elements and determine accordingly fermions of much higher masses.

\begin{table}
\begin{center}
\begin{tabular}{|r||c|c|c|c|}
\hline
$\alpha$&$ I_{R} $&$ II_{R} $&$ III_{R} $&$ IV_{R}$\\
\hline\hline
$I_{L}$   & $ A^I_{\alpha}  $ & $ \tilde{A}^{++}_{\alpha} ((03),(12)) $ 
& $  \tilde{A}^{++}_{\alpha} ((03),(78)) $  &
$ -  \tilde{A}^{++}_{\alpha} ((12),(78)) $ \\
\hline
$II_{L}$  & $ \tilde{A}^{--}_{\alpha} ((03),(12))$ & $ A^{II}_{\alpha} $ & $ 
\tilde{A}^{-+}_{\alpha} ((12),(78)) $ &
$ -  \tilde{A}^{-+}_{\alpha} ((03),(78)) $ \\
\hline 
$III_{L}$ & $  \tilde{A}^{--}_{\alpha} ((03),(78)) $ & 
$- \tilde{A}^{+-}_{\alpha} ((12),(78)) $ & $ A^{III}_{\alpha}$ &
$   \tilde{A}^{-+}_{\alpha} ((03),(12)) $ \\
\hline 
$IV_{L}$  & $ \tilde{A}^{--}_{\alpha} ((12),(78)) $ & $- \tilde{A}^{+-}_{\alpha} ((03),(78))  $ & 
$ \tilde{A}^{+-}_{\alpha} ((03),(12))$ & $ A^{IV}_{\alpha}  $ \\
\hline\hline
\end{tabular}
\end{center}
\caption{\label{TableII.} %
The mass matrix of four families of  $u$-quarks 
obtained within the approach 
unifying spins and charges under the assumptions i.-iii. and a. (in section~\ref{Lwithassumptions}). 
The fields $A^{i}_{\alpha}$, $i=I,II,III,IV$ and $\tilde{A}^{kl}_{\alpha} ((ab),(cd))$,
$k,l=\pm$ and $(ab), (cd) = (03),(12), (78)$ are expressible with the corresponding 
$\tilde{\omega}_{abc\alpha}$ fields (Eq.(\ref{yukawatilde0})).  
They then accordingly determine 
the properties of the  four families of  $u$-quarks. The mass matrix is not yet 
required to be symmetric and real. }
\end{table}

\begin{table}
\begin{center}
\begin{tabular}{|r||c|c|c|c|}
\hline
$\beta$&$ I_{R} $&$ II_{R} $&$ III_{R} $&$ IV_{R}$\\
\hline\hline
$I_{L}$   & $ A^I_{\beta}  $ & $ \tilde{A}^{++}_{\beta} ((03),(12)) $ 
& $ - \tilde{A}^{++}_{\beta} ((03),(78)) $  &
$     \tilde{A}^{++}_{\beta} ((12),(78)) $ \\
\hline
$II_{L}$  & $ \tilde{A}^{--}_{\beta} ((03),(12))$ & $ A^{II}_{\beta} $ & $ 
-  \tilde{A}^{-+}_{\beta} ((12),(78)) $ &
$  \tilde{A}^{-+}_{\beta} ((03),(78)) $ \\
\hline 
$III_{L}$ & $  -\tilde{A}^{--}_{\beta} ((03),(78)) $ & 
$  \tilde{A}^{+-}_{\beta} ((12),(78)) $ & $ A^{III}_{\beta}$ &
$   \tilde{A}^{-+}_{\beta} ((03),(12)) $ \\
\hline 
$IV_{L}$  & $ - \tilde{A}^{--}_{\beta} ((12),(78)) $ & $ \tilde{A}^{+-}_{\beta} ((03),(78))  $ & 
$ \tilde{A}^{+-}_{\beta} ((03),(12))$ & $ A^{IV}_{\beta}  $ \\
\hline\hline
\end{tabular}
\end{center}
\caption{\label{TableIII.}%
The mass matrix of four families of  $d$-quarks 
obtained within the approach 
unifying spins and charges under the assumptions i.-iii. and a. (in section~\ref{Lwithassumptions}). 
Comments are the same as in 
 Table~\ref{TableII.}.} 
\end{table}

If requiring that the mass matrices are real and symmetric, one ends up with the 
matrix elements for  the $u$-quarks as follows: 
$ \tilde{A}^{++}_{\alpha} ((03),(12))= 
 \frac{1}{2}(\tilde{\omega}_{327 \alpha} +\tilde{\omega}_{ 018 \alpha})
= \tilde{A}^{--}_{\alpha} ((03),(12)),$ 
$  \tilde{A}^{++}_{\alpha} ((03),(78)) = 
 \frac{1}{2}(\tilde{\omega}_{ 387 \alpha} +\tilde{\omega}_{078 \alpha}) = 
\tilde{A}^{--}_{\alpha} ((03),(78)),$
$  \tilde{A}^{++}_{\alpha} ((12),(78)) = - 
 \frac{1}{2}(\tilde{\omega}_{277 \alpha} +\tilde{\omega}_{187 \alpha})= 
 - \tilde{A}^{--}_{\alpha} ((12),(78))$, 
$\tilde{A}^{-+}_{\alpha} ((12),(78)) = -
 \frac{1}{2}(\tilde{\omega}_{277 \alpha} -\tilde{\omega}_{187 \alpha})=
 -\tilde{A}^{+-}_{\alpha} ((12),(78))$, 
$ -  \tilde{A}^{-+}_{\alpha} ((03),(78)) = 
 \frac{1}{2}(\tilde{\omega}_{387 \alpha} - \tilde{\omega}_{078 \alpha})=
 \tilde{A}^{+-}_{\alpha} ((03),(78))$, 
$\tilde{A}^{-+}_{\alpha} ((03),(12)) = 
 -\frac{1}{2}(\tilde{\omega}_{327 \alpha} -\tilde{\omega}_{018 \alpha})= 
\tilde{A}^{+-}_{\alpha} ((03),(12)).$ The diagonal terms are 
$A^{II}_{\alpha} = A^{I}_{\alpha} +  
(\tilde{\omega}_{127 \alpha} - \tilde{\omega}_{038 \alpha})$, 
 $A^{III}_{\alpha}= A^{I}_{\alpha} +  
 (\tilde{\omega}_{787 \alpha} - \tilde{\omega}_{038 \alpha})$, 
 $A^{IV}_{\alpha}= A^{I}_{\alpha} + 
 (\tilde{\omega}_{127 \alpha} + \tilde{\omega}_{787 \alpha})$. 
 
 One obtains equivalent expressions also for the $d$-quarks: 
 $ \tilde{A}^{++}_{\beta} ((03),(12))= 
  \frac{1}{2}(\tilde{\omega}_{327 \beta} - \tilde{\omega}_{ 018 \beta})
 = \tilde{A}^{--}_{\beta} ((03),(12)),$ 
 $  \tilde{A}^{++}_{\beta} ((03),(78)) = 
  \frac{1}{2}(\tilde{\omega}_{ 387 \beta} - \tilde{\omega}_{078 \beta}) = 
 \tilde{A}^{--}_{\beta} ((03),(78)),$
 $  \tilde{A}^{++}_{\beta} ((12),(78)) = - 
  \frac{1}{2}(\tilde{\omega}_{277 \beta} +\tilde{\omega}_{187 \beta})= 
  - \tilde{A}^{--}_{\beta} ((12),(78))$, 
 $\tilde{A}^{-+}_{\beta} ((12),(78)) = -
  \frac{1}{2}(\tilde{\omega}_{277 \beta} -\tilde{\omega}_{187 \beta})=
  -\tilde{A}^{+-}_{\beta} ((12),(78))$, 
 $ -  \tilde{A}^{-+}_{\beta} ((03),(78)) = 
 - \frac{1}{2}(\tilde{\omega}_{387 \beta} + \tilde{\omega}_{078 \beta})=
  \tilde{A}^{+-}_{\beta} ((03),(78))$, 
 $\tilde{A}^{-+}_{\beta} ((03),(12)) = 
  -\frac{1}{2}(\tilde{\omega}_{327 \beta} + \tilde{\omega}_{018 \beta})= 
 \tilde{A}^{+-}_{\beta} ((03),(12)).$ 
 The diagonal terms are 
 $A^{II}_{\beta} = A^{I}_{\beta} +  
 (\tilde{\omega}_{127 \beta} + \tilde{\omega}_{038 \beta})$, 
  $A^{III}_{\beta}= A^{I}_{\beta} +  
  (\tilde{\omega}_{787 \beta} + \tilde{\omega}_{038 \beta})$, 
  $A^{IV}_{\beta}= A^{I}_{\beta} + 
  (\tilde{\omega}_{127 \beta} + \tilde{\omega}_{787 \beta})$. 
Different parameters for the members of the families are due to different expressions for the 
matrix elements, different diagonal terms, contributed by $S^{ab}\omega_{ab\pm}$ and also due to 
perturbative and nonperturbative effects which appear through breaks of symmetries.

Let us assume that the mass matrices are real and symmetric (assumption iv. in 
section \ref{Lwithassumptions}) and in addition   
that the break of symmetries leads to  two heavy and two light families and that  
 the mass matrices are diagonalizable in a two steps process \cite{mdn06,matjazdiploma} 
 so that the first diagonalization transforms the mass matrices into block-diagonal 
 matrices with two  $2\times 2$ sub-matrices. We follow the ref.~\cite{mdn06} 
 (where the reader can find all the details).  
It is easy to prove that a $4\times 4$ matrix is diagonalizable in two steps only  
 if it has a  structure 
\begin{equation}\label{deggen}
          \left(\begin{array}{cc}
                  A   &  B\nonumber\\
                  B   &  C=A+k B \nonumber\\
                    \end{array}
                \right).
\end{equation}

Since $A$ and $C$ are assumed to be symmetric  
$2 \times 2$ matrices, 
so must be $B$. 
The parameter $k$, which is an unknown  parameter,  has  
the property that $k= k_u= -k_d$, where the index $u$ or $d$ denotes the $u$ and the $d$ quarks,  
respectively. 
The above assumption  requires that $\tilde{\omega}_{277\delta}=0,
\tilde{\omega}_{377\delta}= - \frac{k}{2} \tilde{\omega}_{187\delta}, 
\tilde{\omega}_{787\delta}=  \frac{k}{2} \tilde{\omega}_{387\delta}, 
\tilde{\omega}_{038\delta}= - \frac{k}{2} \tilde{\omega}_{078\delta}, \delta = u,d. 
$

Under these assumptions the matrices diagonalizing the 
mass matrices are expressible with only three parameters, and the angles of rotations in 
the $u$-quark case are related to the angles of rotations in the $d$-quark case as follows
\begin{eqnarray}
\tan {}^{a,b}\varphi_{u,d} &=& (\sqrt{1+({}^{a,b}\eta_{u,d})^2}\mp {}^{a,b}\eta_{u,d}),
\nonumber\\
{}^{a}\eta_{u} &=& 
- {}^{a}\eta_{d},\quad 
{}^{b}\eta_{u} = 
- {}^{b}\eta_{d}, 
\label{anglesrotex}
\end{eqnarray}
with $a$, which determines the lower two times two matrices and $b$  the higher two times 
two matrices  after the first step diagonalization. Then the angles of rotations in 
the $u$ and the $d$ 
quark case are related: i. For the angle of the first rotation 
(which leads to two by diagonal matrices) we find $\tan \varphi_u = \tan^{-1} \varphi_d $, 
with $\varphi_u = \frac{\pi}{4}- \frac{\varphi}{2}$. 
ii. For the angles of the second rotations in the sector $a$ and $b$ 
we correspondingly find for the $u$-quark ${}^{a,b}\varphi_u = \frac{\pi}{4} 
- \frac{{}^{a,b}\varphi}{2}$ and for the $d$-quark 
${}^{a,b}\varphi_d = \frac{\pi}{4} + \frac{{}^{a,b}\varphi}{2}$. 

It is now easy to express all the fields $\tilde{\omega}_{abc}$ in terms of the masses and 
the parameters $k$ and ${}^{a,b}\eta_{u,d}$ 
\begin{eqnarray}
\label{omegatilde}
\tilde{\omega}_{018 u} &=&\frac{1}{2}[\frac{m_{u 2} - 
m_{u 1}}{\sqrt{1+ ({}^a\eta_{u})^2}} +
                \frac{m_{u 4 } - m_{u 3}}{ \sqrt{1+ ({}^b\eta_{u})^2}}], \nonumber\\
\tilde{\omega}_{078 u} &=&\frac{1}{2 \;\sqrt{1+ (\frac{k}{2})^2}}
[\frac{{}^a\eta_{u}\;(m_{u 2 } - m_{u 1})}{\sqrt{1+ ({}^a\eta_{u})^2}} - 
 \frac{{}^b\eta_{u}\;(m_{u 4 } - m_{u 3})}{\sqrt{1+ ({}^b\eta_{u})^2}}], \nonumber\\
 \tilde{\omega}_{127 u} &=&\frac{1}{2}[\frac{{}^a\eta_{u}\;(m_{u 2} - 
m_{u 1})}{\sqrt{1+ ({}^a\eta_{u})^2}} +
\frac{{}^b\eta_{u}\;(m_{u 4 } - m_{u 3})}{\sqrt{1+ ({}^b\eta_{u})^2}}], \nonumber\\
\tilde{\omega}_{187 u} &=&\frac{1}{2 \sqrt{1+(\frac{k}{2})^2}}[-\frac{
m_{u 2} - m_{u 1}}{\sqrt{1+ ({}^a\eta_{u})^2}} +
\frac{m_{u 4 } - m_{u 3}}{ \sqrt{1+ 
({}^b\eta_{u})^2}}], \nonumber\\
\tilde{\omega}_{387u} &=& \frac{1}{2 \sqrt{1+(\frac{k}{2})^2}}
[(m_{u 4 } + m_{u 3}) - (m_{u 2} + m_{u 1})],\nonumber\\
a_{u}&=& \frac{1}{2}( m_{u 1} + m_{u 2} - \frac{{}^a\eta_{u}\;(m_{u 2} - 
m_{u 1})}{\sqrt{1+({}^a\eta_{u})^2}}),
\end{eqnarray}
with $a_u = A^{I}_u -   \frac{1}{2} \tilde{\omega}_{038u} + 
\frac{1}{2}(\frac{k}{2}- \sqrt{1+(\frac{k}{2})^2 })
(\tilde{\omega}_{078u} + \tilde{\omega}_{387\delta}) 
$,  and equivalently for the $d$-quarks, where ${}^{a,b}\eta_u$ stays unchanged 
(Eq.(\ref{anglesrotex})). 

The experimental data offer the masses of six quarks and the corresponding mixing 
matrix for the three families (within the measured  accuracy and the corresponding 
calculational errors). Due to our assumptions  the mixing  matrix is real and 
antisymmetric 
\begin{eqnarray}
\label{abcwithm}
{\bf V}_{ud} = \left(\begin{array}{cccc}
c(\varphi)c({}^a\varphi)&-c(\varphi)s({}^a\varphi)&
-s(\varphi)c({}^a\varphi^b)& s(\varphi)s({}^a\varphi^b)\\
c(\varphi)s({}^a\varphi)&c(\varphi)c({}^a\varphi)&
-s(\varphi)s({}^a\varphi^b)& -s(\varphi)c({}^a\varphi^b)\\
s(\varphi)c({}^a\varphi^b)&-s(\varphi)s({}^a\varphi^b)&
c(\varphi)c({}^b\varphi)& -c(\varphi)s({}^b\varphi)\\
s(\varphi)s({}^a\varphi^b)&s(\varphi)c({}^a\varphi^b)&
c(\varphi)s({}^b\varphi)& c(\varphi)c({}^a\varphi)\\
 \end{array}
                \right),
\end{eqnarray}
where
\begin{eqnarray}
\label{anglesandmixmatr3ngles}
\varphi = \varphi_{\alpha}-\varphi_{\beta},
\quad {}^a\varphi = {}^a\varphi_{\alpha}-{}^a\varphi_{\beta},\quad 
{}^a\varphi^b = - \frac{{}^a\varphi + {}^b \varphi}{2}, 
\end{eqnarray}
with the angles  described  by the three parameters 
$k,{}^{a}\eta_{u},{}^{b}\eta_{u}$.

We present numerical results in the next section. The assumptions which we made 
left us with the problem of fitting  twelve parameters  
for both types of quarks with the experimental data. 
Since the parameter $k$, which determines the first step of diagonalization of 
mass matrices, turns out (experimentally) to be very small, the ratios of the 
fields $\tilde{\omega}_{abc}$ for $u$-quarks  and $d$-quarks 
($\frac{\tilde{\omega}_{abc\, u}}{\tilde{\omega}_{abc\,d} }$) are almost determined  
with the values ${}^{a,b}\eta_{u}$ (Eq.(\ref{anglesrotex})) and we are left with 
seven parameters, which should be fitted to  twice three masses 
of quarks and (in our simplified case) three angles within the known accuracy.

\subsection{Four families of quarks in proposal no. II}
\label{GN4}

The assumption made in the previous subsection~(\ref{MDN4}) takes care---in the 
$S^{ab}$ sector---that   
the mass term  conserves the electromagnetic charge. The same assumption was made also in   
the $\tilde{S}^{ab}$ sector.

In this subsection we study the break of the symmetries 
from $ SO(1,7)\times U(1)\times SU(3)$ down to 
$SO(1,3) \times  U(1)\times SU(3)$ which occurs in  the following steps. 
First we assume that all the matrix elements  $\tilde{A}^{kl}_{\pm} ((ab),(cd)), $ 
which have   $(ab)$ equal to either $(03)$ or $(12)$ and  $(cd)$  
equal to either $(56)$ or $(78)$ are equal to zero,   which means that the symmetry 
$SO(1,7)\times U(1)$ breaks into $SO(1,3)\times SO(4)\times U(1)$.  

We then break $SO(4)\times U(1)$ in the sector $\tilde{S}^{ab} \tilde{\omega}_{abs}, s=7,8, $ 
so that at some high scale one of $SU(2)$ in $SO(4)\times U(1)$ breaks together with $U(1)$ 
into $SU(2)\times U(1)$ and then---at much lower scale, which is the weak scale---the break 
of the symmetry of $SU(2)\times U(1)$ to $U(1)$ appears.

The break of the symmetries 
from $ SO(1,7)\times U(1)$ to $SO(1,3)\times SO(4)\times U(1)$ makes 
the eight families to decouple into two times four families,  arranged as follows 
\begin{eqnarray}
I.\;& & \stackrel{03}{[+i]} \stackrel{12}{(+)} |\stackrel{56}{(+)} \stackrel{78}{[+]}||...\quad \quad
  V.\; \stackrel{03}{(+i)} \stackrel{12}{(+)} |\stackrel{56}{(+)} \stackrel{78}{(+)}||...\nonumber\\ 
II.\;& &\stackrel{03}{(+i)} \stackrel{12}{[+]} |\stackrel{56}{(+)} \stackrel{78}{[+]}||... \quad \quad
  VI.\;\stackrel{03}{[+i]} \stackrel{12}{[+]} |\stackrel{56}{(+)} \stackrel{78}{(+)}||... \nonumber\\
III.& & \stackrel{03}{[+i]} \stackrel{12}{(+)} |\stackrel{56}{[+]} \stackrel{78}{(+)}||... \quad \quad
  VII. \stackrel{03}{(+i)} \stackrel{12}{(+)} |\stackrel{56}{[+]} \stackrel{78}{[+]}||... \nonumber\\
IV. & & \stackrel{03}{(+i)} \stackrel{12}{[+]} |\stackrel{56}{[+]} \stackrel{78}{(+)}||... \quad \quad
VIII.  \stackrel{03}{[+i]} \stackrel{12}{[+]} |\stackrel{56}{[+]} \stackrel{78}{[+]}||... \;\;.
\label{fourfamilies1}
\end{eqnarray}
We shall see  that the parameters of the second four families lead accordingly 
 to much higher masses.

In Eq.(\ref{yukawatilde0}) we rearranged the terms $\tilde{S}^{ab}\tilde{\omega}_{ab \pm}$  for 
$a,b= 0,1,2,3,5,6,7,8 $ in terms of the raising and lowering operators,  which are 
products of nilpotents $\stackrel{ab}{\tilde{(\pm k_1)}} \stackrel{cd}{\tilde{(\pm k_2)}}$,
with $(ab),(cd)$ belonging to the Cartan subalgebra. Introducing the notation 
for the particular lowering and raising operators as follows 
\begin{eqnarray}
\tilde{\tau}^{+}_{N_+} &=& -\stackrel{03}{\tilde{(-i)}} \stackrel{12}{\tilde{(+)}}, \quad
\tilde{\tau}^{-}_{N_+}  =  -\stackrel{03}{\tilde{(+i)}} \stackrel{12}{\tilde{(-)}}, \quad
\tilde{\tau}^{+}_{N_-}  =   \stackrel{03}{\tilde{(+i)}} \stackrel{12}{\tilde{(+)}}, \quad
\tilde{\tau}^{-}_{N_-}  =  -\stackrel{03}{\tilde{(-i)}} \stackrel{12}{\tilde{(-)}}, \nonumber\\
\tilde{\tau}^{1+}      &=& -\stackrel{56}{\tilde{(+)}}  \stackrel{78}{\tilde{(-)}}, \quad\;
\tilde{\tau}^{1-}  =  \;\;\; \stackrel{56}{\tilde{(-)}}  \stackrel{78}{\tilde{(+)}}, \quad\;\;\,
\tilde{\tau}^{2+}\;\;   =   \stackrel{56}{\tilde{(+)}}  \stackrel{78}{\tilde{(+)}}, \quad
\tilde{\tau}^{2-}\;\,   =  -\stackrel{56}{\tilde{(-)}}  \stackrel{78}{\tilde{(-)}}, 
\label{matrixY2tau}
\end{eqnarray}
and for the diagonal operators  
\begin{eqnarray}
\tilde{N}^{3}_{+} &=& \frac{1}{2}(\tilde{S}^{12} +i \tilde{S}^{03}), \;
\tilde{N}^{3}_{-}  =  \frac{1}{2}(\tilde{S}^{12} -i \tilde{S}^{03}), \;
\tilde{\tau}^{13}  =  \frac{1}{2}(\tilde{S}^{56} -  \tilde{S}^{78}), \;
\tilde{\tau}^{23}  =  \frac{1}{2}(\tilde{S}^{56} +  \tilde{S}^{78}),
\label{matrixY2diag}
\end{eqnarray}
 we can write 
\begin{eqnarray}
& & \frac{1}{2}\tilde{S}^{ab}\tilde{\omega}_{ab \pm} = 
    \frac{\tilde{g}^m}{\sqrt{2}} 
(-\tilde{\tau}^{+}_{N_+} \tilde{A}^{+ N_+}_{\pm} - \tilde{\tau}^{-}_{N_+} \tilde{A}^{- N_+}_{\pm} +
 \tilde{\tau}^{+}_{N_-} \tilde{A}^{+ N_-}_{\pm} + \tilde{\tau}^{-}_{N_-} \tilde{A}^{- N_-}_{\pm}) 
\nonumber\\
&+ &\frac{\tilde{g}^1}{\sqrt{2}} 
(-\tilde{\tau}^{1+}      \tilde{A}^{1+}_{\pm}  +   \tilde{\tau}^{1-}  \tilde{A}^{1-}_{\pm}) +
\frac{\tilde{g}^2}{\sqrt{2}}
( \tilde{\tau}^{2+}      \tilde{A}^{2+}_{\pm}  +   \tilde{\tau}^{2-}  \tilde{A}^{2-}_{\pm}) 
\nonumber\\ 
& +&\tilde{g}^m 
      (\tilde{N}^{3}_{+} \tilde{A}^{3 N_+}_{\pm} + \tilde{N}^{3}_{-} \tilde{A}^{3 N_-}_{\pm} + 
     \tilde{g}^1 
(\tilde{\tau}^{13} \tilde{A}^{13}_{\pm}    +       \tilde{\tau}^{23} \tilde{A}^{23}_{\pm} )  
\nonumber\\
&+& \tilde{g}^{4} \tilde{\tau}^{4} \tilde{A}^{4}_{\pm}.
\label{matrixY2}
\end{eqnarray}

The fields  $\tilde{A}^{kl}_{\pm}((ab) (cd))$ in Eq.(\ref{Awithomega}) and the 
fields in Eq.(\ref{matrixY2}) if we take them together with the coupling constants 
$\tilde{g}^i, i=1,2,4,m$, (taking care of the running in the $\tilde{S}^{ab}$ sector) 
are in one to one correspondence. 
For example,  
$ -\frac{\tilde{g}^m}{\sqrt{2}} \tilde{\tau}^{+}_{N_+} \tilde{A}^{+ N_+}_{\pm}= 
-\stackrel{03}{\tilde{(-i)}}  \stackrel{12}{\tilde{(+)}}\tilde{A}^{-+}_{\pm}. $ 

We assume that at the  break of $SO(4)\times U(1)$ into $SU(2)\times U(1)$ appearing at some 
large scale  new fields  $\tilde{A}^{Y}_{\pm}$  and $\tilde{A}^{Y'}_{\pm}$ manifest 
(in a similar way  in the Standard model new fields occur when the weak charge breaks)
\begin{eqnarray}
\label{newfields} 
\tilde{A}^{23}_{\pm} &=& \tilde{A}^{Y}_{\pm} \sin \tilde{\theta}_2 + 
\tilde{A}^{Y'}_{\pm} \cos \tilde{\theta}_2, \nonumber\\
\tilde{A}^{41}_{\pm} &=& \tilde{A}^{Y}_{\pm} \cos \tilde{\theta}_2 -  
\tilde{A}^{Y'}_{\pm} \sin \tilde{\theta}_2
\end{eqnarray}
and accordingly also new operators 
\begin{eqnarray}
\label{newoperators}
\tilde{Y}&=& \tilde{\tau}^{41}+ \tilde{\tau}^{23}, \quad 
\tilde{Y'}= \tilde{\tau}^{23} - \tilde{\tau}^{41} \tan \tilde{\theta}_2.
\end{eqnarray}
It  then follows for the $\tilde{S}^{ab}\tilde{\omega}_{ab \pm}$ sector  of the mass matrix  
\begin{eqnarray}
& &\frac{1}{2}\tilde{S}^{ab}\tilde{\omega}_{ab \pm} = \frac{\tilde{g}^m}{\sqrt{2}} 
(-\tilde{\tau}^{+}_{N_+} \tilde{A}^{+ N_+}_{\pm} - \tilde{\tau}^{-}_{N_+} \tilde{A}^{- N_+}_{\pm} +
\tilde{\tau}^{+}_{N_-} \tilde{A}^{+ N_-}_{\pm} + \tilde{\tau}^{-}_{N_-} \tilde{A}^{- N_-}_{\pm}) +
\nonumber\\
& &\frac{\tilde{g}^1}{\sqrt{2}} 
(-\tilde{\tau}^{1+} \tilde{A}^{1+}_{\pm} +   \tilde{\tau}^{1-} \tilde{A}^{1-}_{\pm}) +
\frac{\tilde{g}^2}{\sqrt{2}}
(\tilde{\tau}^{2+} \tilde{A}^{2+}_{\pm}  +   \tilde{\tau}^{2-} \tilde{A}^{2-}_{\pm}) +
\nonumber\\
& & \tilde{g}^m  
(\tilde{N}^{3}_{+} \tilde{A}^{3 N_+}_{\pm} + \tilde{N}^{3}_{-} \tilde{A}^{3 N_-}_{\pm} + 
\tilde{g}^Y \tilde{A}^{Y}_{\pm} \tilde{Y} + \tilde{g}^{Y'}  \tilde{A}^{Y'}_{\pm} \tilde{Y'}
+ \tilde{\tau}^{13} \tilde{A}^{13}_{\pm} ).
\label{matrixY2a}
\end{eqnarray}
Here $\tilde{g}^Y= \tilde{g}^{4} \cos \tilde{\theta}_2$,  
$ \tilde{g}^{Y'} = \tilde{g}^2 \cos \tilde{\theta}_2$ and $\tan \tilde{\theta}_2 = 
\frac{\tilde{g}^4}{\tilde{g}^2} $. 

Let at the weak scale the $SU(2)\times U(1)$ break further into $U(1)$  leading again 
to new fields
\begin{eqnarray}
\label{newfieldsweak}
\tilde{A}^{13}_{\pm} &=& \tilde{A}_{\pm} \sin \tilde{\theta}_1 + 
\tilde{Z}_{\pm} \cos \tilde{\theta}_1,\nonumber\\ 
\tilde{A}^{Y}_{\pm} &=& \tilde{A}_{\pm} \cos \tilde{\theta}_1 -  
\tilde{Z}_{\pm} \sin \tilde{\theta}_1
\end{eqnarray}
and new operators 
\begin{eqnarray}
\label{newoperatorsweak}
\tilde{Q}  &=&  \tilde{\tau}^{13}+ \tilde{Y} = \tilde{S}^{56} +  \tilde{\tau}^{41},\nonumber\\
\tilde{Q'} &=& -\tilde{Y} \tan^2 \tilde{\theta}_1 + \tilde{\tau}^{13},
\end{eqnarray}
with $\tilde{e} = 
\tilde{g}^{Y}\cos \tilde{\theta}_1, \tilde{g'} = 
\tilde{g}^{1}\cos \tilde{\theta}_1$  and $\tan \tilde{\theta}_1 = 
\frac{\tilde{g}^{Y}}{\tilde{g}^1} $. 
If $\tilde{\theta}_2 $ appears to be  very small and 
$\tilde{g}^2 \tilde{A}^{2\pm}_{\pm}$ and $\tilde{g}^{Y'}  \tilde{A}^{Y'}_{\pm} \tilde{Y'}$ 
very large, the second four families (decoupled from the first one) appear to be very 
heavy in comparison with the first four families. The first four families 
mass matrix (evaluated on a tree level) for  the $u$-quarks ($-$) and the $d$-quarks ($+$) 
is presented in Table~\ref{TableIV.}. 
\begin{table}
\begin{center}
\renewcommand{\arraystretch}{1.5}
\begin{tabular}{||c|c|c|c|c||}
\hline
$ $ & $I$  & $ II$ & $ III $ & $ IV $ \\
\hline\hline
$ I$  & $ a_{\pm}$ & $ \frac{\tilde{g}^m}{\sqrt{2}} \tilde{A}^{+ N_+ }_{\pm}
$
 & $-\frac{\tilde{g}^1}{\sqrt{2}} \tilde{A}^{1+}_{\pm} $ & $0$ \\
\hline
$II$ & $ \frac{\tilde{g}^m}{\sqrt{2}}\tilde{A}^{- N_+ }_{\pm}
 $
& $a_{\pm} + \frac{1}{2}\tilde{g}^m(\tilde{A}^{3 N_-}_{\pm} + \tilde{A}^{3 N_+}_{\pm})$
   & $0$
   & $
   -\frac{\tilde{g}^1}{\sqrt{2}}\tilde{A}^{1+}_{\pm} $ \\
\hline
$III$ & $\frac{\tilde{g}^1}{\sqrt{2}} \tilde{A}^{1-}_{\pm} $
 & $ 0$
 & $ a_{\pm} +  \tilde{e} \tilde{A}_{\pm} + \tilde{g'} \tilde{Z}_{\pm} $
 & $ \frac{\tilde{g}^m}{\sqrt{2}}\tilde{A}^{+N_+}_{\pm} $\\
\hline
$IV$ & $ 0$
 & $ \frac{\tilde{g}^1}{\sqrt{2}}\tilde{A}^{1-}_{\pm} $
 & $ \frac{\tilde{g}^m}{\sqrt{2}}\tilde{A}^{-N_+}_{\pm} $
 & $ a_{\pm} + \tilde{e} \tilde{A}_{\pm} + \tilde{g'} \tilde{Z}_{\pm} + \frac{1}{2} \tilde{g}^m
  (\tilde{A}^{3 N_-}_{\pm}+ \tilde{A}^{3 N_+}_{\pm})$\\
\hline\hline
\end{tabular}
\end{center}
\caption{\label{TableIV.}%
The mass matrix for the lower four families of the $u$-quarks (with the sign $-$) and 
the $d$-quarks (with the sign $+$).}
\end{table}

In Table~\ref{TableIV.} 
$a_{\pm}$ stands for the contribution to the mass matrices from the $S^{ab}\omega_{ab\pm}$ part 
(which distinguishes among the members of each particular family) and from the diagonal terms of the 
$\tilde{S}^{ab}\tilde{\omega}_{ab\pm}$ part. The mass matrix in Table~\ref{TableIV.}
is in general complex. To be able to make  an estimate of the properties of the four families 
of quarks we assume (as in subsection~\ref{MDN4}) that  the mass matrices 
are real and symmetric. We then treat the elements as they appear in  Table~\ref{TableIV.} as 
free parameters and fit them to the experimental data. Accordingly we rewrite 
the mass matrix in Table~\ref{TableIV.} in the form presented in Table~\ref{TableV.}. 

\begin{table}
\begin{center}
\renewcommand{\arraystretch}{0.5}
\begin{tabular}{|c|cccc|}
\hline
$ $   & $I$             & $ II$                 & $ III $                 & $ IV $ \\
\hline
$ I$  & $ a_{\pm} $     & $ b_{\pm}$            & $-c_{\pm} $             & $0$    \\
$II$  & $ b_{\pm} $     & $ a_{\pm} + d_{1\pm}$ & $0$                     & $-c_{\pm} $ \\
$III$ & $c_{\pm}  $     & $ 0            $      & $ a_{\pm} +  d_{2\pm} $ & $ b_{\pm} $\\
$IV$  & $ 0$            & $ c_{\pm} $           & $ b_{\pm} $  & $ a_{\pm} + d_{3\pm} $\\
\hline
\end{tabular}
\end{center}
\caption{\label{TableV.}%
The mass matrix from Table~\ref{TableIV.},  taken in this case to be real  
and parameterized in a transparent way. $-$, $+$ denote the $u$-quarks and the $d$-quarks, 
respectively. }
\end{table}

The parameters $b_{\pm}$, $c_{\pm}$, $d_{i\pm}, i=1,2,3$ are expressible in terms
of the real and symmetric part of the matrix elements of Table~\ref{TableIV.}.   
We present the way of adjusting  parameters 
to the experimental data for  the three known families in the next section.


\section{Numerical results}
\label{numerical}


The two types of mass matrices in section~\ref{Lwithassumptions}  
followed from the two assumed ways of breaking symmetries 
from $SO(1,7)\times U(1)\times SU(3)$ down to the observable 
$SO(1,3)\times U(1)\times SU(3)$ in the scalar (with respect to $SO(1,3)$) part 
determining the Yukawa couplings. Since the problem of deriving the Yukawa couplings explicitly 
from the starting Lagrange density of the approach unifying 
spins and charges is  very complex, we make in this paper   a rough 
estimation for each of the two proposed breaks of symmetries in order to see whether 
the approach can be the right way to go  beyond the Standard model of the electroweak and 
colour interactions and what does the approach teach us about the families. We hope that the 
perturbative and nonperturbative effects manifest at least to some extent  
in the parameters of the mass 
matrices, which we leave to be adjusted so that the masses and the mixing matrix 
for the three known families  of quarks agree (within the declaired accuracy) 
with the experimental data. We also investigate a possibility of making   
predictions about the properties 
of the fourth family.

\subsection{Experimental data for quarks}
\label{experimentaldata}

We take  the experimental data   for the known three 
families of quarks from ref.~\cite{expckm,expmixleptons}. We use for masses the data 
\begin{eqnarray}
\label{masses}
m_{u_i}/\textrm{GeV} &=& (0.0015- 0.005, 1.15-1.35, 174.3-178.1
),\nonumber\\
m_{d_i}/\textrm{GeV} &=& (0.004-0.008, 0.08-0.13, 4.1-4.9 
).
\end{eqnarray}
Predicting four families of quarks and leptons at ''physical'' energies, we require 
the unitarity condition for the mixing matrices for four rather than three measured families 
of quarks~\cite{expckm}  
\begin{eqnarray}
\label{expckm}
 \left(\begin{array}{ccc}
 0.9730-0.9746 & 0.2157-0.22781 & 0.0032-0.0044\\
 0.220-0.241   & 0.968-0.975  & 0.039-0.044\\
 0.006 - 0.008    & 0.035-0.043      & 0.07-0.9993\\
\end{array}
                \right).\nonumber\\
|V_{td}/V_{ts}|=0.208 ^{+0.008}_{-0.006}  \;  \textrm{or}  \;  0.16\pm0.04.
\end{eqnarray}
We keep in mind that the ratio of the mixing matrix elements 
$|V_{td}/V_{ts}|$ includes the assumption that there exist only three families.  


\subsection{Results for proposal no. I}
\label{numericalresults1}


We see that within the experimental accuracy the (real part of the) mixing matrix  
may be assumed to be approximately symmetric up to a sign and then accordingly parameterized 
with only three parameters.
Eq.(\ref{omegatilde}) offers for the  way of breaking the symmetry 
$SO(1,7)\times U(1)\times SU(3)$ down to the observable 
$SO(1,3)\times U(1)\times SU(3)$ proposed in subsection 
(\ref{MDN4}) the relations among the proposed elements of the two mass matrices for quarks 
on one and the masses of quarks and  the three angles 
determining the mixing matrix on the other side.  
We have $7$ parameters to be fitted to the six measured  masses  
and  the measured elements of the  mixing matrix within the experimental 
accuracy.
We use the Monte-Carlo method  to adjust the parameters to the experimental data presented 
in Eqs.(\ref{masses},\ref{expckm}). 
We allow the two quark masses of the fourth  
family to lie in the range from $200\,$GeV to $1\,$TeV. 
The obtained results  for $k$ and the two ${}^{a,b}\eta$ are presented in Table~\ref{TableVI.}. 
\begin{table}
\centering
\begin{tabular}{|c||c|c|c|c|}
\hline 
&$u$&$d$\tabularnewline
\hline
\hline 
$k$       & -0.085 &  0.085\tabularnewline
\hline 
${}^a\eta$& -0.229 &  0.229\tabularnewline
\hline
${}^b\eta$&  0.420 & -0.440\tabularnewline
\hline
\end{tabular}\\
\caption{\label{TableVI.}%
The Monte-Carlo fit to the experimental data~\cite{expckm,expmixleptons} 
for the parameters $k$, ${}^a\eta$ and   ${}^b\eta$ determining the mixing matrices 
for  the four families of quarks is presented.}
\end{table}
In Table~\ref{TableVII.} the fields  $\tilde{\omega}_{abc}$   are 
presented.  
\begin{table}
\begin{tabular}{|c||c|c|c||c|c|c|}
\hline 
&$u$&$d$&$u/d$\tabularnewline
\hline
\hline 
$|\tilde{\omega}_{018}|$& 21205& 42547& 0.498\tabularnewline
\hline 
$|\tilde{\omega}_{078}|$& 49536& 101042& 0.490\tabularnewline
\hline 
$|\tilde{\omega}_{127}|$& 50700& 101239& 0.501\tabularnewline
\hline 
$|\tilde{\omega}_{187}|$& 20930& 42485& 0.493\tabularnewline
\hline 
$|\tilde{\omega}_{387}|$& 230055& 114042& 2.017\tabularnewline
\hline
$a$&94174& 6237& \tabularnewline
\hline
\end{tabular}\\
\caption{\label{TableVII.}%
Values for the parameters $\tilde{\omega}_{abc }$ in MeV  
for the $u-$quarks and the $d-$quarks (subsection~\ref{MDN4}) 
as obtained by the Monte-Carlo fit  
relating the parameters and the experimental data.}
\end{table}
One notices that the Monte-Carlo fit keeps the ratios of the 
$\tilde{\omega}_{abc } $ very close to $0.5$ ($k$ is small but not zero). 
In Eq.(\ref{resultmassesM}) we present the corresponding masses for the four families of 
quarks  
\begin{eqnarray}
\label{resultmassesM}
m_{u_i}/\textrm{GeV} &=& (0.0034, 1.15, 176.5, 285.2),\nonumber\\
m_{d_i}/\textrm{GeV} &=& (0.0046, 0.11, 4.4, 224.0),
\end{eqnarray}
and the mixing matrix for the quarks
\begin{eqnarray}
\label{resultckmM}
 \left(\begin{array}{cccc}
 0.974 & 0.223 & 0.004 & 0.042\\
 0.223 & 0.974 & 0.042 & 0.004\\
 0.004 & 0.042 & 0.921 & 0.387\\
 0.042 & 0.004 & 0.387 & 0.921\\
 \end{array}
                \right).
\end{eqnarray}
For the ratio $|V_{td}/V_{ts}|$ we find in Eq.(\ref{resultckmM})  
the value arround $~0.1$. 
The estimated mixing matrix for the four families of quarks predicts quite a strong couplings  
between the fourth and the other three families, limiting some of the matrix elements 
of the three families as well.


\subsection{Results for proposal no. II}
\label{numericalresults2}


In  subsection~\ref{GN4} assumptions about the  way of breaking the symmetries 
(from $ SO(1,7)\times U(1)\times SU(3)$  to the "physical" ones  
 $SO(1,3) \times U(1)\times SU(3)$) leave us with 
two four families of very different masses for the $u$ and the $d$ quarks. 
In Table~\ref{TableV.} the mass matrices for the lighter of the two four families of quarks 
are presented in a parameterized way under the  assumption that the mass matrices 
are real and symmetric. 

There are six free parameters in each of the two mass matrices. 
The  two off diagonal elements together with three out of  four diagonal elements 
determine the orthogonal transformation, which diagonalizes the mass matrix
(subtraction of a constant times the unit matrix does not change the orthogonal 
transformation). 
The four times four matrix is diagonizable with the orthogonal 
transformation depending on six angles (in general with $n(n-1)/2$).  
We use the Monte-Carlo method to fit the free parameters of each 
of the two mass matrices to the elements of the quark 
mixing matrix Eqs.(\ref{expckm})  and the quark masses Eqs.(\ref{masses}) of the 
known three families. 
One  notices that the matrix in Table~\ref{TableV.} splits into two times two matrices, 
if we put parameters $c_{\pm}$ equal to zero. Due to experimental data we expect that $c_{\pm}$ 
must be small. The quark mixing matrix is assumed to be real (but not also 
symmetric as it was in~\ref{numericalresults1}).  
Since  there are more free parameters than the experimental data to be fitted, we 
look for the best fit in dependence on the quark masses of the fourth family.
Assuming for the fourth family quark masses the values $m_{u_4}= 285\,$GeV and 
$m_{u_d}= 215\,$GeV  
the Monte-Carlo fit gives the following mass matrices (in MeV) 
($(-b,-a)\cup(a,b)$ means that both intervals are taken into account)
for the $u$-quarks 
\begin{eqnarray}
\label{mu}
\left(
\begin{array}{cccc}
(9,22) & {(-150,-83)\atop\cup(83,150)} & (-50,50) & (-306,304) \\
{(-150,-83)\atop\cup(83,150)} & (1211,1245) & (-306, 304) & (-50,50) \\
(-50,50) & (-306,304) & (171600,176400) & {(-150,-83)\atop\cup(83,150)} \\
(-306, 304) & (-50,50) & {(-150,-83)\atop\cup(83,150)} & (200000, 285000) \\
\end{array}
\right)
\end{eqnarray}
and for the $d$-quarks  
\begin{eqnarray}
\label{md}
\left(
\begin{array}{cccc}
(5,11)& {(8.2,14.5)\atop\cup (-14.5,-8.2)}& (-50,50) & {(-198,-174)\atop\cup(174,198)}  \\
{(8.2,14.5)\atop\cup (-14.5,-8.2)}& (83 - 115) & {(-198,-174)\atop\cup(174,198)}& (-50,50) \\
(-50,50) & {(-198,-174)\atop\cup(174,198)}  & (4260 - 4660) & {(8.2,14.5)\atop\cup (-14.5,-8.2)} \\
{(-198,-174)\atop\cup(174,198)} & (-50,50) & {(8.2,14.5)\atop\cup (-14.5,-8.2)} & (200000,215000) \\
\end{array}
\right).
\end{eqnarray}
The above mass matrices correspond to the following values 
for the quark masses (the central values are written only) 
\begin{eqnarray}
\label{resultmassesG}
m_{u_i}/\textrm{GeV} &=& (0.005, 1.220, 171., 285.),\nonumber\\
m_{d_i}/\textrm{GeV} &=& (0.008, 0.100, 4.500, 215.),  
\nonumber
\end{eqnarray}

\noindent
and to the following absolute values for the quark mixing matrix 
(the central values are written only)
\begin{eqnarray}
\label{resultckmG}
 \left(\begin{array}{cccc}
	0.974&0.226&0.00412&0.00218 \\
	0.226&0.973&0.0421&0.000207 \\
	0.0055&0.0419&0.999&0.00294 \\
	0.00215&0.000414&0.00293&0.999 
 \end{array}
                \right)
 \end{eqnarray}
 with 80 $\%$ 
 confidence level. 
We get $|V_{td}|/|V_{ts}|= 0.128 - 0.149$.

For higher values of the two masses of the fourth family the matrix elements 
of the mixing matrix $V_{i4}$ and $V_{4i}, i=d,s,t$, are slowly decreasing---decoupling very slowly 
the fourth families from the first three. For $m_{u_4}= 500 {\rm GeV}= m_{d_4}$, for example, 
we obtain
$V_{d4} < 0.00093$, $V_{s4} < 0.00013$, $0.00028 < V_{b4} < 0.00048$, 
$V_{4u} < 0.00093$, $V_{4c} < 0.00015$, $0.00028 < V_{4t} < 0.00048$. 



\section{Discussions and conclusions}
\label{discussions}

We study in this paper 
whether the approach of one of 
us~\cite{norma92,norma93,normasuper94,norma95,pikanormaproceedings1,holgernorma00,norma01,%
pikanormaproceedings2,Portoroz03} unifying spins and charges 
might  answer those of the open questions of the Standard model of the 
electroweak and colour interactions which are connected with the appearance of  
families of fermions, of the Yukawa couplings and of  the weak scale. 

Starting from the Lagrange density for spinors in $d(=1+13)$-dimensional 
space with two kinds of  fields (Eq.(\ref{lagrange}))---the 
gauge fields ($\omega_{abc}$) of the Lorentz group ($S^{ab}= \frac{i}{4}
(\gamma^a\gamma^b- \gamma^b\gamma^a)$) and the gauge fields ($\tilde{\omega}_{abc}$) of 
$\tilde{S}^{ab}$ ($= \frac{i}{4}
(\tilde{\gamma}^a\tilde{\gamma}^b - \tilde{\gamma}^b \tilde{\gamma}^a),$ 
with  $\tilde{\gamma}^a$ which anticommute with $\gamma^a$, $ \{\gamma^a,
\tilde{\gamma}^b\}_+ =0$)---we end up, by assuming two possible breaks of symmetries 
(subsections~\ref{MDN4},\ref{GN4}), at   
the "physical" scale with two types of four families of quarks and leptons corresponding to 
the two chosen ways of breaking symmetries. 
In this paper we study only the properties of quarks. 
Parameterization of the quark masses depends strongly on the chosen way of breaking symmetries.

One Weyl spinor in $d=1+13$ if analyzed in terms of the Standard model groups $SO(1,3) 
\times SU(2)\times U(1)\times SU(3)$ manifests the left handed weak charged quarks and leptons 
and weak chargeless right handed quarks and leptons. 
It is a long way from the starting simple Lagrange density for spinors carrying only the spins 
and interacting with only  the vielbeins and the spin connections  (two kinds of)
(Eq.(\ref{lagrange})) to the observable quarks and leptons. 
To treat breaking of the starting symmetries properly, taking into account 
all perturbative and nonperturbative effects, boundary conditions and other effects  
(by treating gauge gravitational fields in the same way as ordinary gauge fields, 
since the scale of breaking $SO(1,13)$ is supposed to be far from the Planck scale) 
is a huge project.

In this paper we estimate the part $\psi^{\dagger} 
\gamma^0 \gamma^s p_{0s} \psi,$ with $s=7,8,$ of the starting Lagrange density, which 
determines the Yukawa couplings (Eq.(\ref{yukawa},\ref{yukawatilde0})) and does 
what in the Standard model the Higgs field is doing. 
The two chosen ways of breaking the starting symmetries (subsections~\ref{MDN4},\ref{GN4}) 
in the $\tilde{S}^{ab}$ sector differ in the part breaking   
$SO(1,7)\times U(1)$ with eight massless families  
to $SO(1,3)\times U(1)$ where four families have low enough masses to be all four 
possibly observeble at ''physical energies'', while the rest four families 
with much higher masses  are decoupled from the lower four families.  

In one of the two ways of breaking symmetries we assume that   there are no 
matrix elements  of the type $\tilde{A}^{k l}_{\pm} ((ab),(cd))$, 
with $k=\pm $ and $l=\pm $ (in all four combinations) and with either 
$(ab)$ or $(cd)$  equal to $(56)$. (In the Poincar\' e sector such a choice guarantees 
the conservation of the electromagnetic charge.) We also assume that the mass matrices  
are symmetric and real (hoping that this assumption does not influence considerably the 
real part of the mixing matrices and the masses)  and diagonalizable in two steps. 

In the second  choice of breaking the starting symmetries we instead 
assume that all the matrix elements  $\tilde{A}^{kl}_{\pm} ((ab),(cd)), $  
which have  $(ab)$ equal to either $(03)$ or $(12)$ and  $(cd)$  
equal to $(56)$ or $(78)$ are equal to zero,   
which means that the symmetry $SO(1,7)\times U(1)$ breaks into 
$SO(1,3)\times SO(4)\times U(1)$.   
We then break $SO(4)\times U(1)$ in the sector $\tilde{\omega}_{abs}, s=7,8, $ 
so that at some high scale one of $SU(2)$ in $SO(4)\times U(1)$ breaks together with $U(1)$ 
into $SU(2)\times U(1)$ and then---at much lower scale---the break 
of the symmetry of $SU(2)\times U(1)$ to $U(1)$ appears. We again end up 
with the four families decoupled from the much heavier four families with 
the quark mass matrices differing strongly from 
the mass matrices in the first choice. We assume again that the 
mass matrices are 
real.

We make the calculations on the tree level, obtaining mass matrices for quarks 
in both chosen ways  parameterized 
with the fields, whose strengths depend on the way and the scale of breaking 
symmetries. We let the perturbative and nonperturbative effects to be (at least to some 
extent) 
included in the  fields, for which we assume that they are free parameters to be determined 
by fitting the masses and the mixing matrix to the known experimental data 
within the known accuracy. 

The symmetries of the mass matrices in the first chosen way of breaking the starting symmetries 
locate (after assuming the real and symmetric mass matrices, diagonalizable in two steps)
the masses of the four families to be in the 
region for which the analyzes in refs.~\cite{okun,okunmaltoni,okunbulatov} show that it 
is experimentally allowed.  The second choice of breaking the symmetries (although each of the 
mass matrices having only two off diagonal elements) does not determine the masses  
of the fourth quark family, leaving these masses as free parameters.
Both choices predict rather strong coupling among the observed three and the fourth family. 
The fourth family decouples in the second choice of breaking symmetries 
from the first three for pretty high values for the fourth family quark 
masses. The calculated ratio $|V_{td}|/|V_{ts}|$ 
differs  for both assumed ways of breaking the symmetries from the measured one 
(we obtain $|V_{td}|/|V_{ts}|$ equal to 
$0.128 - 0.149$ in the second case), which is expected   
since the measured value is obtained with the inclusion of the 
calculations made under the assumption 
that there are only three families.  

Both chosen ways are very approximate. To say which of the two ways is more trustable, 
further (more demanding) calculations have to be made. However,  it seems quite acceptable that 
breaks of symmetries go in both sectors---the Poincar\' e one and the one 
connected  with $\tilde{S}^{ab}$---through two steps of breaking the symmetries 
$SO(1,7)\times U(1)$ (as suggested by  
the second way of breaking symmetries) resembling in both steps the 
Standard model way of breaking the symmetry in the spinor sector, suggesting that 
the second way might be 
the right one, although one can not at all expect that the break of symmetries 
in both sectors manifests in the same way. This paper is to be  understood as a 
first step to further calculations, which should at the end tell whether our way of describing 
charges and families is the right way beyond the Standard model 
of the electroweak and colour interactions.

Let us make a note that the decoupled four families might be candidates for forming the 
dark matter~\cite{khlopovproc}. 

\section*{Acknowledgments} We would like to express many thanks to ARRS for the
grant.
It is a pleasure to thank all the participants of the   workshops entitled 
"What comes beyond the Standard model", 
taking place  at Bled annually in  July, starting at 1998,  for fruitful discussions, 
in particular to H.B. Nielsen.


\end{document}